\documentclass{article}

\usepackage{arxiv}

\usepackage[utf8]{inputenc} 
\usepackage[T1]{fontenc}    
\usepackage{hyperref}       
\usepackage{url}            
\usepackage{booktabs}       
\usepackage{amsfonts}       
\usepackage{nicefrac}       
\usepackage{microtype}      
\usepackage{graphicx}
\usepackage{natbib}
\usepackage{doi}
\usepackage{bm}
\usepackage{amsmath}
\usepackage{multirow}
\usepackage{subcaption}

\title{Dynamic Skewness in Stochastic Volatility Models: A Penalized Prior Approach}

\author{
        Bruno E. Holtz\thanks{Instituto de Ciências Matemáticas e de Computação, Universidade de São Paulo. Full address: Av. Trabalhador São-carlense, 400 – Centro, 13566-590, São Carlos – SP, Brazil.}\\
        bruno.holtz@usp.br\\
        \and
        Ricardo S. Ehlers \footnotemark[1]\\
        ehlers@icmc.usp.br\\
        \and
        Adriano K. Suzuki\footnotemark[1]\\
        suzuki@icmc.usp.br\\
        \and
        Francisco Louzada\footnotemark[1]\\
        louzada@icmc.usp.br
}

\hypersetup{
            pdftitle={A template for the arxiv style},
            pdfsubject={q-bio.NC, q-bio.QM},
            pdfauthor={David S.~Hippocampus, Elias D.~Striatum},
            pdfkeywords={First keyword, Second keyword, More},
}

\begin{document}
\maketitle

\begin{abstract}
	Financial time series often exhibit skewness and heavy tails, making it essential to use models that incorporate these characteristics to ensure greater reliability in the results. Furthermore, allowing temporal variation in the skewness parameter can bring significant gains in the analysis of this type of series. However, for more robustness, it is crucial to develop models that balance flexibility and parsimony. In this paper, we propose dynamic skewness stochastic volatility models in the SMSN family (DynSSV-SMSN), using priors that penalize model complexity. Parameter estimation was carried out using the Hamiltonian Monte Carlo (HMC) method via the \texttt{RStan} package. Simulation results demonstrated that penalizing priors present superior performance in several scenarios compared to the classical choices. In the empirical application to returns of cryptocurrencies, models with heavy tails and dynamic skewness provided a better fit to the data according to the DIC, WAIC, and LOO-CV information criteria.
\end{abstract}


\section{Introduction}
\label{sec-intro}

Stochastic volatility (SV) models are a crucial tool for analyzing financial time series. These models are flexible enough to capture the nonlinear behavior usually observed in financial time series of returns. In their original formulation, \cite{taylor1986} assumes that observational errors of the model follow a standard normal distribution. However, various empirical studies indicate that this formulation is unrealistic, and several authors have proposed extensions to the basic SV model to incorporate the distinctive characteristics of financial time series. For example, \cite{abanto2010robust} extend the SV model by allowing observational errors to belong to the Scale Mixture of Normals (SMN) family (see also \cite{wang-etal}). When studying the leverage effect in SV models, \cite{huang2014stochastic} propose an extension that allows observed returns to exhibit heavy tails and skewness simultaneously, using the family of Scale Mixture of Skew Normal (SMSN) distribution. 

More recent studies consider models with asymmetric observational errors, where the degree of skewness varies over time. \cite{iseringhausen2020time} investigates this issue by analyzing exchange rates, assuming that the errors follow a non-centered Student's $t$ distribution. In their studies, they found that the degree of skewness exhibits significant variations over time and argued that this model can enhance the accuracy of Value-at-Risk (VaR) forecasting. However, the prior distributions adopted to capture the dynamics of skewness do not allow this parameter to be constant, which can lead to overfitting. In contrast, \cite{martins2024stochastic} use penalized prior distributions in the process of generating skewness, allowing the model to be symmetric or to have dynamic skewness. 

When it comes to prior specification in SV models, it is common practice to adopt prior distributions that have already been used before in similar applications. These traditional priors are intended to be vaguely informative and belong to standard families of distributions. However, priors that penalize the complexity of financial models with variable parameters have attracted the interest of researchers in recent years. For example, \cite{bitto2019achieving} use a double gamma distribution as a prior in the variance process, allowing the model to be automatically reduced to the homoscedastic case. Empirical results indicate that this class of priors is effective in avoiding overfitting when the coefficients governing the dynamics of the variance are static or non-significant. In the context of State Space Models, \cite{lopes2022parsimony} proposed a mixture prior that enables the investigation of different types of state evolution within the structure of a first-order autoregressive process.

Several recent studies have applied GARCH-type models to Bitcoin and other cryptocurrencies, reflecting a growing interest in this type of asset. For instance, \cite{ding2024bitcoin}, \cite{sozen2025volatility}, \cite{mostafa2021gjr}, and \cite{wu2025cryptocurrency}. However, there is a noticeable scarcity of research that employs SV models in this setting. A notable exception is \cite{kim2021forecasting}, who compare a Bayesian SV model with standard GARCH across nine major cryptocurrencies (including Bitcoin, Ethereum and XRP) and report that the SV model delivers superior volatility forecasts, particularly in periods of extreme market fluctuations. Despite this contribution, the literature integrating stochastic volatility and dynamic asymmetry, particularly within the flexible SMSN family, is still limited in applications to cryptocurrency markets, to the best of our knowledge.

The main contribution of this paper to these lines of research is threefold. The first approach is to expand the traditional SV model by incorporating possible heavy-tailedness and skewness simultaneously, using a scale mixture of skew-normal distributions. This further flexibility is important, as there is evidence in the financial literature that the distribution of financial returns often exhibits fat tails and occasional asymmetry. In addition, even greater flexibility is achieved by allowing the degree of skewness to vary over time. The second contribution is to employ the Penalized Complexity Prior (PCP), as proposed by \cite{simpson2017penalising}, to control the degree of skewness in the stochastic volatility model and then compare it with the priors adopted by \cite{iseringhausen2020time} and \cite{martins2024stochastic}. The PCP offers a principled way to favour simpler models, unless the data strongly supports additional complexity. By embedding the parsimony principle directly into the prior specification, the model can balance flexibility and interpretability while reducing the risk of overfitting. Our simulation results show that penalized priors outperform the classical one in several scenarios, with the PCP yielding the best overall performance. The third contribution is empirical: by applying the proposed model to daily cryptocurrency returns, we find evidence of dynamic asymmetry that oscillates smoothly around zero over time, indicating that the conditional distribution of these returns tends to alternate between slightly negative and slightly positive asymmetry.

The remainder of this paper is organized as follows. Section~\ref{section:Section2} describes the SMSN family, the DynSSV-SMSN class model, and the parameter estimation using the Bayesian paradigm. The theoretical foundations and construction of the PCP adopted in this paper are discussed in detail in Section~\ref{section:Section3}. Section~\ref{section:Section4} develops simulation studies. Section ~\ref{section:Section5} is intended for empirical application. Finally, some conclusions and suggestions for future developments are discussed in Section \ref{section:Section6}.

\section{A Dynamic Skewness Stochastic Volatility model}
\label{section:Section2}

In the following, we first present the scale mixture of skew normal distributions Family.
A random variable $X$ follows a skew normal distribution with location $\gamma \in \mathbb{R}$, scale $\omega \in \mathbb{R}_{+}$ and skewness parameter $\alpha \in \mathbb{R}$, if it has density function and stochastic representation, which are given \citep{huang2014stochastic} respectively by
\begin{equation*}
    f(x) = \frac{2}{\omega} \phi \left( \frac{(x-\gamma)}{\omega} \right) \Phi \left( \frac{\alpha(x-\gamma)}{\omega} \right)
\end{equation*}
and
\begin{equation}
    X = \gamma + \omega \delta W + \omega \sqrt{1-\delta^{2}} \epsilon,
    \label{sn}
\end{equation}
where $W \sim \mathcal{N}_{[0,\infty)}(0,1)$ and $\epsilon \sim \mathcal{N}(0,1)$ are independent, $\delta\in(-1,1)$ and the skewness parameter $\alpha$ is defined as $\delta = \alpha / \sqrt{1 + \alpha^{2}}$. In this case, we denote $X \sim SN(\gamma, \omega^{2}, \alpha)$. Clearly, for $\alpha = 0$, we have $X \sim \mathcal{N}(\gamma,\omega^{2})$.

Now, let $Z$ be a random variable defined by
\begin{equation}
    Z = \gamma + U^{-1/2}X,
    \label{smsn}
\end{equation}
where $X \sim SN(0, \omega^{2}, \alpha)$ and $U$ is a positive random variable with a cumulative distribution function $F(u|\nu)$, where $\nu>0$ is the parameter that controls the behavior of the tails. In this case, we say $Z \sim SMSN(\gamma, \omega^{2}, \alpha, \nu)$. 
\begin{equation*}
    f(z; \gamma, \omega^{2}) = 2 \int_{0}^{\infty} \psi(z; \gamma, u^{-1}\omega^{2}) \Phi \left( \frac{u^{1/2}\alpha(x-\gamma)}{\omega} \right) dF(u|\nu),
\end{equation*}
where $\psi(.;\mu, \sigma^{2})$ denotes the normal density function with mean $\mu$ and variance $\sigma^{2}$. From (\ref{sn}) and (\ref{smsn}), we can show (see online Appendix A of the supplementary material) that 
\begin{equation}
    E[Z] = \gamma + \sqrt{ \frac{2}{\pi} } \omega \delta k_{1}
    \label{media_smsn}
\end{equation}
and
\begin{equation}
    Var[Z] = \sigma^{2} \gamma,
    \label{var_smsn}
\end{equation}
where $\gamma = \left( k_{2} - \frac{2}{\pi}\delta^{2}k_{1}^{2} \right)$ and $k_{m} = E[ U^{-m/2}]$. Using equation (\ref{var_smsn}) we note that the quantity $\gamma$ needs to be positive, so the variance $Var[Z]$ is well defined. Online Appendix B of the supplementary material shows the conditions necessary for the models studied to satisfy this condition.

As can be seen in equation (\ref{smsn}), the distribution of $Z$ depends on the specification of $U$. For example, when $U \sim \mathcal{G}(\nu / 2, \nu / 2)$, $Z$ follows a t-skew distribution with $\nu$ degrees of freedom and $\alpha$ skewness. When $U = 1$, we have $Z \sim SN(\mu, \sigma^{2}, \alpha)$. In addition to these cases, if $U \sim \mathcal{B}e(\nu, 1)$, $Z$ follows the skewed version of the slash distribution (see \citet{huang2014stochastic}).

\subsection{The Model Specification}

The Dynamic Skewness Stochastic Volatility class model with Scale Mixture of Skew Normal (DynSSV-SMSN) is defined as,
\begin{eqnarray}
    y_{t} &=& \exp(h_{t}/2) \, z_{t}, \nonumber \\ 
    h_{t} &=& \mu + \phi(h_{t-1} - \mu) + \sigma_{h} \epsilon^{h}_{t}, \nonumber \\ 
    \alpha_{t} &= & \alpha_{t-1} + \sigma_{\alpha}\epsilon^{\alpha}_{t} \nonumber \\
    z_{t} &\sim& SMSN(\gamma, \omega^{2}, \alpha_{t}, U_{t}), \label{eq4}
\end{eqnarray}
where $y_{t}$ represents the compound return on day $t$, considering the closing price $P_{t}$. Innovations $\epsilon^{h}_{t}$ and $\epsilon^{a}_{t}$ are mutually and serially independent standard normals. Logarithmic volatility $h_{t}$ follows a stationary AR(1) process with $|\phi|<1$, where the initial state $h_{1} \sim \mathcal{N}(\mu_ {h}, \sigma_{h}^{2}/(1-\phi^{2}))$. $U_{t}$ is the mixing variable with density $p( \cdot | \nu)$, where $\nu$ controls the heavy tail. In order to provide greater flexibility for the model's skewness, the observational innovations $z_{t}$ belong to the SMSN family with the skewness parameter varying over time according to a random walk process staring at $\alpha_{1}$. In order to induce sparsity in the model, we adopt $\alpha_{1}|\kappa \sim Laplace(0, 1/\kappa)$, with $\kappa \sim \mathcal{G}(0.1,0.1)$, as the approach described by \citet{martins2024stochastic}. 

Following \citet{abanto2015bayesian}, we select the quantities $\gamma$ and $\omega$ of $z_{t}$ such that $E[z_{t}] = 0$ and $Var[z_{t}]=1 $, so that the martingale hypothesis is guaranteed. In this way, by equations (\ref{media_smsn}) and (\ref{var_smsn}), we have the following,
\begin{equation*}
    \gamma_{t} = - \sqrt{\frac{2}{\pi}} \omega_{t} \delta_{t} k_{1} \quad \text{and} \quad \omega_{t}^{ 2} = \left(k_{2} - \frac{2}{\pi} \delta_{t}^{2}k_{1}^{2} \right)^{-1}.
\end{equation*}
As described in the beginning of Section \ref{section:Section2}, this class of models includes models with t-Student (DynSSV-t) and slash (DynSSV-S) errors, in addition to the normal case (DynSSV-N). 

In particular, different particular models can be obtained from the values assumed by the parameters $\alpha_{1}$ and $\sigma_{\alpha}$. When $\alpha_{1} = \sigma_{\alpha} = 0$, the model reduces to the symmetric SV-SMN case, as introduced by \citet{abanto2010robust}. On the other hand, if $\alpha_{1} \neq 0$ and $\sigma_{\alpha} = 0$, we obtain the SV-SMSN model, which incorporates skewness in a static way, as proposed by \citet{huang2014stochastic}. Thus, the DynSSV-SMSN class model allows for simultaneous estimation and partial model selection.

The models defined by equations (\ref{eq4}) can be represented hierarchically by the equations (\ref{sn}) and (\ref{smsn}), as
\begin{equation*}
    y_{t} = (\gamma_{t} + \omega_{t} \delta_{t} W_{t} U^{-1/2}_{t}) \exp(h_{t}/2) + U^{-1/2}_{t}\omega_{t}\sqrt{1-\delta^{2}_{t}}\exp(h_{t}/2) \, \epsilon_{t},
\end{equation*}
with $W_{t} \sim \mathcal{N}_{[0, \infty)}(0,1)$, $\epsilon_{t} \sim N(0,1)$ and $\delta_{t} = \alpha_{t} / \sqrt{1 + \alpha_{t}^{2}}$, for $t=1, \ldots, T$. In this way, we can note that
\begin{equation}
    y_{t} | h_{t}, U_{t}, W_{t}, \alpha_{t} \sim \mathcal{N}( \mu_{t}, \sigma^{2}_{t} ),
    \label{hierarchy}
\end{equation}
where
\begin{equation*}
    \mu_{t} = (\gamma_{t} + \omega_{t} \delta_{t} W_{t} U^{-1/2}_{t}) \exp(h_{t}/2)
\end{equation*}
and
\begin{equation*}
    \sigma_{t} = U^{-1/2}_{t}\omega_{t}\sqrt{1-\delta^{2}_{t}}\exp(h_{t}/2).
\end{equation*}
This representation will be useful for the next section.

\subsection{Parameter estimation}

Let $\bm{\theta} = ( \mu, \phi, \sigma_{h}, \alpha_{1}, \kappa, \sigma_{a},  \nu)'$ be the parameter vector for the model class, DynSSV-SMSN, $\bm{h}_{T} = (h_{1}, \ldots, h_{T})'$ the log-volatility vector, $\bm{U}_{T} = (U_{1}, \ldots, U_{T})'$ the mixing variables, $\bm{\alpha}_{T} = (\alpha_{2} , \ldots, \alpha_{T})'$ the skewness coefficients, $\bm{W}_{T} = (W_{1}, \ldots, W_{T})$ and $\bm{y}_{T} = (y_{1}, \ldots, y_{T})$ the information available up to time $T$. 
The model estimation will be carried out according to the Bayesian approach using the principle of data augmentation, which considers $\bm{h}_{T}$, $\bm{U}_{T }$, $\bm{W}_{T}$, $\bm{\alpha}_{T}$ and $\bm{y}_{T}$ as latent variables. The joint posterior density of these variables can be written as
\begin{eqnarray}
    p( \bm{h}_{T}, \bm{U}_{T}, \bm{W}_{T}, \bm{\alpha}_{T}, \bm{\theta} | \bm{y}_{T} ) & \propto& p( \bm{y}_{T} | \bm{h}_{T}, \bm{U}_{T}, \bm{W}_{T}, \bm{\alpha}_{T}, \bm{\theta} ) p(\bm{h}_{T} | \bm{\theta} ) p(\bm{U}_{T} | \bm{\theta}) \times \nonumber \\
    & \times& p(\bm{W}_{T}) p( \bm{\alpha}_{T} | \bm{\theta} ) p(\bm{\theta})\nonumber \\
    &=& \prod_{t=1}^{T} p( {y}_{t} | {h}_{t}, {U}_{t}, {W}_{t}, {\alpha}_{t} ) \prod_{t=1}^{T} p(U_{t}|\bm{\theta}) \nonumber \\
    &\times& p(h_{1}|\bm{\theta}) \prod_{t=2}^{T} p({h}_{t} | h_{t-1}, \bm{\theta} ) \prod_{t=1}^{T} p(W_{t}) \nonumber \\
    &\times& p(\alpha_{1}|\bm{\theta}) \prod_{t=2}^{T} p(\alpha_{t} | \alpha_{t-1}, \bm{\theta} )p(\bm{\theta})
    \label{post}
\end{eqnarray}
where $p( {y}_{t} | {h}_{t}, {U}_{t}, {W}_{t}, {\alpha}_{t})$ is defined by (\ref{hierarchy}) and $p(\bm{\theta})$ is the prior distribution. As the posterior distribution (\ref{post}) has no known form, we used the NUTS method \citep{hoffman2014no}, implemented in the \texttt{rstan} package, available in the \texttt{R} software, to perform all the Bayesian analysis of this work. 

Finally, to complete the model specification, it is necessary to define the prior distribution $p( \bm{\theta})$. In this way, we assume an independent prior,
\begin{equation*}
    p(\bm{\theta}) = p(\mu)p(\phi)p(\sigma_{h}) p(\sigma_{\alpha}) p(\alpha_{1}, \kappa) p(\nu), \nonumber
\end{equation*}
where $\mu \sim \mathcal{N}(0, 10)$, $\frac{\phi + 1}{2} \sim \mathcal{B}e(20, 1.5)$ and $\sigma_{h}^{2} \sim \mathcal{IG}(2.5, 0.025)$. These priors are often found in the literature; see, for instance, \citet{jacquier1994bayesian, chan2017stochastic, abanto2010robust}. The prior for $\nu$ depends on the specification of the distribution of the mixing variable $\bm{U}_{T}$: we adopt vague priors $\nu \sim \mathcal{G}(2.0, 0.1)$ for the DynSSV-t model, and $\nu \sim \mathcal{G}(0.08,0.04)$ for DynSSV-S. The prior specification for $\sigma_{\alpha}$ is discussed in Section~\ref{section:Section3}.

\section{Penalized Complexity Prior}
\label{section:Section3}

Recently, \cite{simpson2017penalising} introduced a new class of priors known as Penalized Complexity Priors (PCP). This prior can accommodate varying levels of information, from vague to informative. Furthermore, the PCP has a natural connection with the Jeffreys prior. Thus, it can be considered a hybrid case between objective and informative priors. In this section, we give a brief description of PCPs and refer the reader to \cite{simpson2017penalising} for a more detailed review of its theoretical and practical aspects.

The PCP was developed on the basis of four elementary principles, all of which assume the existence of a base model. The first is Occam’s Razor, or the parsimony principle, which states that the simplest model is preferable over complex models when there is no support to justify the latter. The second principle is a complexity measure based on the Kullback-Leibler divergence. This principle defines a quantity that measures the amount of information lost when approximating a complex model $f(x|\xi)$ with a simpler model $f(x|\xi = 0)$. The Kullback-Leibler divergence is defined as \begin{equation} D_{KL}( f(\xi) || f_{0} ) = \int_{ x \in \chi } f(x|\xi)\log\left( \frac{f(x|\xi)}{f(x|\xi=0)} \right) dx, 
\label{kld} \end{equation} 
where $f(\xi) := f(x|\xi)$ is the flexible model, $f_{0} := f(x|\xi=0)$ is the base model, such that $\xi$ is referred to as the flexibility parameter. Thus, using quantity (\ref{kld}) to measure the complexity between two models with densities $f$ and $f_{0}$, we can define $d( \xi ) = \sqrt{2D_{KL}( f(\xi) || f_{0} )}$.

The third concept is the constant rate penalization, which penalizes deviations from the base model with a decay rate $r$, that is, \begin{equation} 
\frac{\pi_{d}(d+\delta)}{\pi_{d}(d)} = r^{\delta}, \quad d, \delta > 0, \nonumber 
\end{equation} 
for some constant $0 < r < 1$. This property ensures that the relative change of the prior does not depend on the distance $d$, which is a reasonable choice when there is no additional information. This assumption implies that $\pi(d) = \lambda \exp(-\lambda d)$, for $r = \exp(-\lambda)$. Thus, the prior in the scale of $\xi$ is given by: 
\begin{equation} \pi(\xi) = \lambda \exp( -\lambda d(\xi)) \left| \frac{\partial d(\xi)}{\partial \xi} \right|. \nonumber 
\end{equation}

The fourth and final principle is that absolute ignorance is not practical. That is, to some extent, the researcher has information about the parameter of interest $\xi$. This information can be translated through the scale $\lambda$, which controls the tails of the distribution of $\pi(\xi)$. This can be achieved by considering the quantile $U$ defined by the researcher and solving 
\begin{equation} 
\mathbb{P}( Q(\xi) > U ) = p ,
\label{lambda_choice} 
\end{equation} 
for $\lambda$, where $Q(\xi)$ is an interpretable function and $p$ is the probability mass of the chosen extreme event.

The PCP for the parameter $\sigma_{\alpha}$ in model (\ref{eq4}) has a known form, namely $\sigma_{\alpha} \sim \text{Exp}(\lambda)$ \cite{simpson2017penalising}. So that, using the equation (\ref{lambda_choice}), the hyperparameter $\lambda$ can be defined as,
\begin{equation}
    \lambda = - \frac{\log(p)}{U},
    \label{lambda2}
\end{equation}
where $(U, p) \in \mathbb{R}_{+} \times (0, 1)$ is a user-specified pair of interpretable values. However, selecting the appropriate values for $(U, p)$ in the unobservable structure of the model (\ref{eq4}) is not a trivial task. In this regard, we follow \cite{simpson2017penalising} who argue that the PCP is generally robust to the choice of $\lambda$, unless it is set to an extremely poor value. Specifically, allocating an excessive probability mass near the base model leads to an undue penalization of the flexible model, i.e., taking the pair $(U,p)$ too close to origin $(0, 0)$.

\section{Simulation Study}
\label{section:Section4}

To evaluate the performance of the PCP in estimating the parameters of the DynSSV-t model, we performed a Monte Carlo simulation study, comparing it with the $Exp(1/\kappa)$ prior, where $\kappa \sim \mathcal{G}(0.1, 0.1)$ (Exp), and the $\mathcal{IG}(2.5, 0.025)$ prior ($\mathcal{IG}$), as used by \cite{martins2024stochastic} and \cite{iseringhausen2020time}, respectively. For the PCP, the hyperparameter $\lambda$ was specified using equation (\ref{lambda2}) with $(U, p) = (0.5, 0.5)$, corresponding to a weakly informative prior. 

We generate $m = 300$ replicates with $T = 1500$ observations, considering different scenarios. The parameters $(\mu, \nu)$ were fixed at $(0.0, 8.0)$, and the pair $(\phi, \sigma_{h}) \in \{ (0.95, 0.15), (0.99, 0.10)\}$, while $\sigma_{\alpha} \in \{ 0.01, 0.05, 0.10 \}$, in order to represent low, medium and high variability in the dynamics of skewness. For each simulated dataset, we performed 7000 MCMC iterations, discarding the first 5000 as burn-in, and retaining 2000 samples for posterior analysis.

\begin{figure}[h!]
    \centering
    \includegraphics[width=0.6\textwidth]{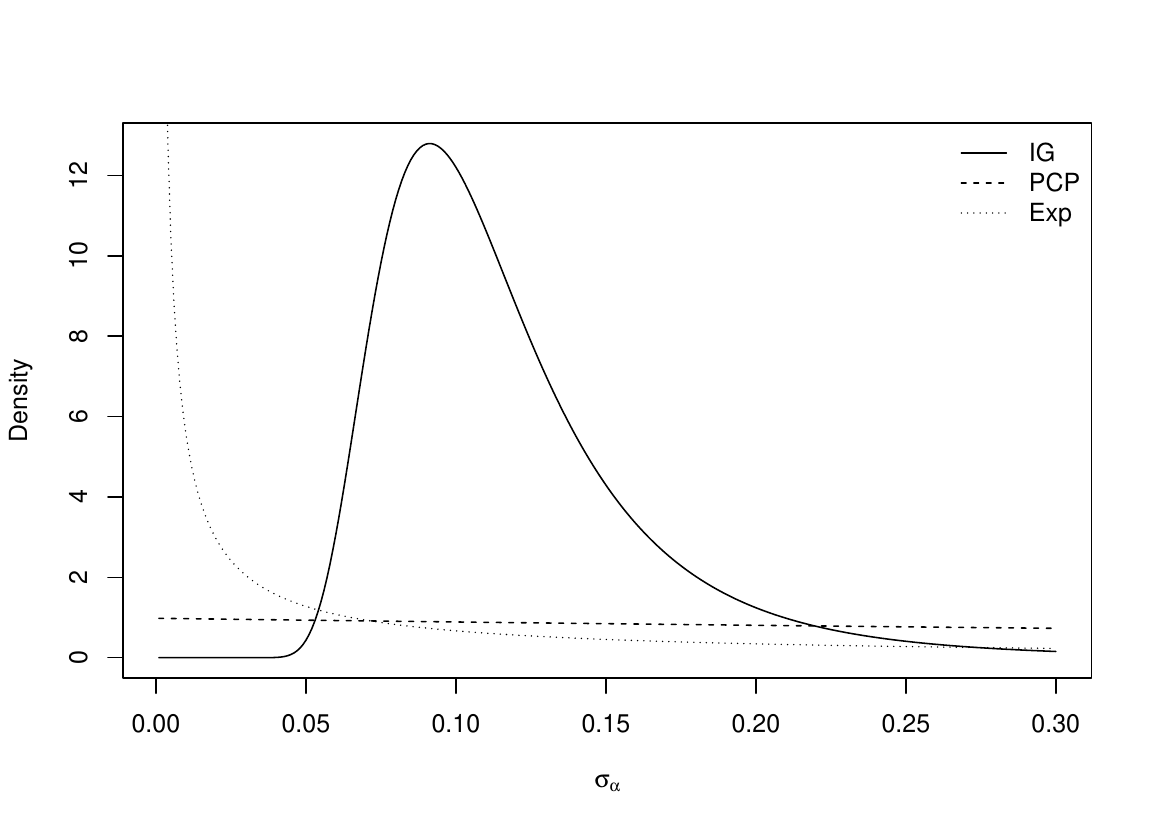}
    \caption{Prior distributions for $\sigma_{\alpha}$ compared in this study.}
    \label{prioris}
\end{figure}
In Figure \ref{prioris}, we compare the density plots of the prior distributions considered. To analyze the impact of the hyperparameters, the figure shows the marginal distribution of the Exp after integration $\kappa$. Notably, the $\mathcal{IG}$ prior assigns a low probability mass to values close to zero. This behavior can be problematic, as it increases the complexity of the model and the possibility of overfitting, as discussed by \cite{simpson2017penalising}. In contrast, the Exp prior places the highest mass near the base model ($\sigma_{\alpha}=0$), thus imposing the strongest penalization. We also note that the PCP prior exhibits a slow tail decay, resembling the behavior of a uniform distribution, which reflects its weakly informative nature. Therefore, we note that the priors have distinct characteristics, with PCP being able to control complexity by allocating mass close to the base model, while allowing $\sigma_{\alpha}$ to assume positive values.

To evaluate the performance of the priors, the relative bias ($\text{Bias}_{rel}$) and the square root of the relative mean square error ($\text{RMSE}_{rel}$) were considered. In addition to these quantities, the probability of coverage was also calculated for the $95\%$ credibility interval with the highest posterior density interval (HPD). Assuming that the proportion of credible intervals containing the true value follows a normal distribution with mean $p = 0.95$ and variance $p(1 - p)/m$, the observed proportions are expected to lie between 0.93 and 0.97 with probability $95\%$. This range serves as a reference for evaluating the coverage performance.
\begin{table}[h!]
    \caption{Results of $\hat{\bm{\theta}}$ estimates of $\mathcal{IG}$ model for different values of $\sigma_{\alpha}$ based on $m=300$ replicates of $T=1500$. True value $\mu=0.0$ and $\nu=8.0$.}
    \centering
    \scalebox{1.0}{
    \begin{tabular}{l c r r r r r r r}
    \toprule
    $\sigma_{\alpha}$ &$\bm{\theta}$ & Mean & Inf & Sup & CD & $\text{Bias}_{rel}$ & $\text{RMSE}_{rel}$ & Coverage \\  
    \midrule
    &\multicolumn{8}{c}{$(\phi,\sigma_{h})=(0.95, 0.15)$} \\
    \cmidrule(l){2-9}
    \multirow{5}{*}{0.01}   &$\mu$              &0.009  &-0.183  &0.203  &0.09   &0.009  &0.096   &0.92 \\
                                                            &$\phi$             &0.946  &0.898  &0.984  &-0.03  &-0.004 &0.023   &0.96 \\
                                                            &$\sigma_{h}$       &0.136  &0.077  &0.205  &-0.07  &-0.093 &0.202   &0.95 \\
                                                            &$\sigma_{\alpha}$  &0.074  &0.047  &0.106  &-0.03  &6.423  &6.437   &0.00 \\
                                                            &$\nu$              &10.501 &5.736  &17.458 &-0.12  &0.313  &0.521   &0.95 \\
                                    \addlinespace
    \multirow{5}{*}{0.05}   &$\mu$              &0.002  &-0.196  &0.201  &-0.01  &0.002  &0.088   &0.96 \\
                                                            &$\phi$             &0.949  &0.904  &0.985  &0.02   &-0.002 &0.019   &0.96 \\
                                                            &$\sigma_{h}$       &0.136  &0.078  &0.204  &-0.04  &-0.091 &0.201   &0.91 \\
                                                            &$\sigma_{\alpha}$  &0.083  &0.051  &0.120  &0.03   &0.657  &0.675   &0.39 \\
                                                            &$\nu$              &10.255 &5.854  &16.660 &0.02   &0.282  &0.537   &0.94 \\
                                    \addlinespace
    \multirow{5}{*}{0.10}   &$\mu$              &0.007  &-0.185  &0.200  &-0.05  &0.007  &0.091   &0.95 \\
                                                            &$\phi$             &0.948  &0.904  &0.984  &-0.03  &-0.002 &0.018   &0.98 \\
                                                            &$\sigma_{h}$       &0.138  &0.080  &0.205  &0.04   &-0.079 &0.187   &0.95 \\
                                                            &$\sigma_{\alpha}$  &0.103  &0.061  &0.154  &-0.10  &0.029  &0.176   &1.00 \\
                                                            &$\nu$              &9.649  &5.959  &14.726 &-0.04  &0.206  &0.443   &0.96 \\
    \cmidrule(l){2-9}
    &\multicolumn{8}{c}{$(\phi,\sigma_{h})=(0.99, 0.10)$} \\
    \cmidrule(l){2-9}
    \multirow{5}{*}{0.01}     &$\mu$    &0.003  &-0.580  &0.579  &0.03   &0.003  &0.245   &0.94 \\
                                                        &$\phi$              &0.985  &0.972  &0.997  &0.06   &-0.005 &0.009   &0.98 \\
                                                        &$\sigma_{h}$        &0.108  &0.073  &0.148  &-0.12  &0.082  &0.169   &0.99 \\
                                                        &$\sigma_{\alpha}$   &0.070  &0.046  &0.098  &0.03   &5.983  &6.008   &0.00 \\
                                                        &$\nu$               &10.732 &5.868  &17.578 &-0.06  &0.342  &0.565   &0.94 \\
    \addlinespace
    \multirow{5}{*}{0.05}   &$\mu$               &0.031  &-0.490  &0.551  &-0.02  &0.031  &0.245   &0.93 \\
                            &$\phi$              &0.985  &0.971  &0.997  &0.01   &-0.006 &0.009   &0.97 \\
                            &$\sigma_{h}$        &0.107  &0.072  &0.146  &-0.07  &0.070  &0.170   &0.98 \\
                            &$\sigma_{\alpha}$   &0.080  &0.051  &0.114  &0.06   &0.598  &0.620   &0.47 \\
                            &$\nu$               &10.450 &5.989  &16.735 &-0.02  &0.306  &0.498   &0.95 \\    
    \addlinespace
    \multirow{5}{*}{0.10}   &$\mu$              &0.090  &-0.430  &0.610  &-0.04  &0.090  &0.295   &0.91 \\
                            &$\phi$              &0.984  &0.970  &0.996  &-0.05  &-0.006 &0.010   &0.97 \\
                            &$\sigma_{h}$        &0.109  &0.074  &0.148  &-0.01  &0.087  &0.175   &0.99 \\
                            &$\sigma_{\alpha}$   &0.101  &0.061  &0.149  &0.03   &0.012  &0.159   &0.99 \\
                            &$\nu$               &9.812  &6.061  &15.089 &0.00   &0.227  &0.428   &0.95 \\
    \bottomrule
    \end{tabular}
    \label{tab1}
    }
\end{table}
In addition to the metrics mentioned above, we also consider the posterior means of the parameters, the lower (Inf) and upper (Sup) limits with respect to the estimated HPD intervals, and the \cite{geweke1991evaluating} statistics (CD), in which values between -1.96 and 1.96 indicate convergence of the chain. Tables $\ref{tab1}$, $\ref{tab2}$ and $\ref{tab3}$ summarize the results obtained. 

In general, we observed good results for the parameters $\mu$, $\phi$, $\sigma_{h}$ and $\nu$, which presented relative bias and RMSE close to zero in all models and scenarios considered, with the exception of $\nu$ for the prior Exp when $\sigma_{\alpha}=0.10$ and $(\phi, \sigma_{h}) = (0.99, 0.1)$. In this case, the relative bias and RMSE exceeded 1.0, and the average HPD interval presented the largest amplitude for $\nu$, indicating a possible inadequacy of the prior in this specific scenario. Furthermore, we can note from the tables that the diagnostics \cite{geweke1991evaluating} (CD) show the convergence of all chains.

The parameter of interest $\sigma_{\alpha}$, however, presented a very different behavior for the models under study. From Table $\ref{tab1}$ we can see that, in both scenarios of $(\phi, \sigma_{h})$, the bias and the relative RSME decay to zero as the true value of $\sigma_{\alpha}$ increases. Interestingly, when $\sigma_{\alpha}=0.01$, the estimated means are of the order of $0.07$, indicating that the prior $\mathcal{IG}$ generates a strong bias that induces complexity in the model, which in turn can generate overfitting. This behavior becomes even more evident when we analyze the lower limits of the HPD intervals, which are significantly above the true value of $\sigma_{\alpha}$, reflecting the low coverage probabilities, as well as the high bias and RMSE. A similar pattern is observed when $\sigma_{\alpha} = 0.05$, although with a relative bias and RMSE closer to zero. On the other hand, in scenarios with greater dynamics in skewness ($\sigma_{\alpha} = 0.10$), the same prior presents the best performance in terms of the metrics considered, when compared to the others. Thus, we can conclude that the prior $\mathcal{IG}$ induces complexity in a systematic way, with more pronounced effects when the variability of $\sigma_{\alpha}$ is low.

\begin{table}[h!]
    \caption{Results of $\hat{\bm{\theta}}$ estimates of $Exp$ model for different values of $\sigma_{\alpha}$ based on $m=300$ replicates of $T=1500$. True value $\mu=0.0$ and $\nu=8.0$.}
    \centering
    \scalebox{1.0}{
    \begin{tabular}{l c rrrrrrr}
    \toprule
    $\sigma_{\alpha}$ &$\bm{\theta}$ & Mean & Inf & Sup & CD & $\text{Bias}_{rel}$ & $\text{RMSE}_{rel}$ & Coverage \\   
    \midrule
    &\multicolumn{8}{c}{$(\phi,\sigma_{h})=(0.95, 0.15)$} \\
    \cmidrule(l){2-9}
    \multirow{5}{*}{0.01}   &$\mu$               &0.007  &-0.185  &0.201  &-0.05 &0.007  &0.090  &0.96 \\
                            &$\phi$              &0.948  &0.904   &0.984  &-0.03 &-0.002 &0.021  &0.97 \\
                            &$\sigma_{h}$        &0.138  &0.078   &0.206  &0.02  &-0.081 &0.210  &0.93 \\
                            &$\sigma_{\alpha}$   &0.003  &0.000   &0.011  &-0.01 &-0.719 &0.883  &0.31 \\
                            &$\nu$               &9.884  &5.607   &15.947 &0.02  &0.236  &0.501  &0.92 \\
    \addlinespace
    \multirow{5}{*}{0.05}   &$\mu$               &0.011  &-0.188  &0.223  &-0.03  &0.011   &0.103  &0.95 \\
                            &$\phi$              &0.947  &0.903   &0.984  &-0.06  &-0.003  &0.021  &0.95 \\
                            &$\sigma_{h}$        &0.139  &0.080   &0.207  &0.08   &-0.075  &0.197  &0.96 \\
                            &$\sigma_{\alpha}$   &0.041  &0.013   &0.078  &-0.03  &-0.182  &0.732  &0.60 \\
                            &$\nu$               &9.651  &5.756   &14.885 &0.04   &0.206   &0.404  &0.95 \\

    \addlinespace
    \multirow{5}{*}{0.10}   &$\mu$               &-0.004  &-0.195  &0.189  &-0.03 &-0.004  &0.088  &0.96 \\
                            &$\phi$              &0.949  &0.907   &0.984  &-0.02 &-0.001  &0.018  &0.97 \\
                            &$\sigma_{h}$        &0.137  &0.080   &0.202  &0.00  &-0.087  &0.198  &0.94 \\
                            &$\sigma_{\alpha}$   &0.105  &0.045   &0.179  &0.05  &0.052   &0.491  &0.85 \\
                            &$\nu$               &9.924  &6.127   &15.276 &0.00  &0.241   &0.426  &0.94 \\
    \cmidrule(l){2-9}
    &\multicolumn{8}{c}{$(\phi,\sigma_{h})=(0.99, 0.10)$} \\
    \cmidrule(l){2-9}
    \multirow{5}{*}{0.01}   &$\mu$               &0.007     &-0.525    &0.545     &0.02     &0.007     &0.253   &0.92 \\
                            &$\phi$              &0.985     &0.972     &0.997     &0.08     &-0.005    &0.008   &0.97 \\
                            &$\sigma_{h}$        &0.109     &0.074     &0.148     &-0.13    &0.087     &0.172   &0.98 \\
                            &$\sigma_{\alpha}$   &0.004     &0.000     &0.013     &-0.08    &-0.633    &0.886   &0.35 \\
                            &$\nu$               &10.536    &6.008     &16.782    &-0.10    &0.317     &0.529   &0.91 \\
    \addlinespace
    \multirow{5}{*}{0.05}   &$\mu$               &0.016     &-0.540    &0.564     &-0.10    &0.016     &0.239   &0.95 \\
                            &$\phi$              &0.985     &0.971     &0.996     &-0.14    &-0.006    &0.010   &0.96 \\
                            &$\sigma_{h}$        &0.110     &0.075     &0.149     &0.10     &0.097     &0.184   &0.97 \\
                            &$\sigma_{\alpha}$   &0.044     &0.015     &0.080     &-0.09    &-0.126    &0.558   &0.76 \\
                            &$\nu$               &10.158    &6.020     &15.918    &-0.04    &0.270     &0.477   &0.93 \\
    \addlinespace
    \multirow{5}{*}{0.10}   &$\mu$               &-0.054    &-0.669    &0.560     &0.01     &-0.054    &0.575   &0.69 \\
                            &$\phi$              &0.898     &0.739     &0.997     &0.10     &-0.092    &0.103   &0.96 \\
                            &$\sigma_{h}$        &0.114     &0.057     &0.189     &-0.04    &0.137     &0.226   &0.99 \\
                            &$\sigma_{\alpha}$   &0.043     &0.002     &0.156     &-0.04    &-0.565    &1.002   &0.44 \\
                            &$\nu$               &18.168    &3.979     &40.633    &-0.08    &1.271     &1.467   &0.98 \\
    \bottomrule
    \end{tabular}
    \label{tab2}
    }
\end{table}
Table \ref{tab2} shows the results for the Exp prior. Once again, the estimates of the parameter $\sigma_{\alpha}$ exhibit a markedly different behavior. In scenarios with lower values of $\sigma_{\alpha}$, particularly when $\sigma_{\alpha}=0.01$, the estimates are strongly biased downward and present low coverage probabilities. This suggests that the Exp prior tends to excessively penalize the model in the presence of small values of $\sigma_{\alpha}$. As the true value of $\sigma_{\alpha}$ increases, both the relative bias and the RMSE decrease, while the coverage improves, indicating better estimation performance. This systematic improvement with increasing $\sigma_{\alpha}$ mirrors the behavior observed under the $\mathcal{IG}$ prior, although with milder biases in intermediate regimes. Nevertheless, under extreme persistence and reduced volatility -- that is, in the scenario $(\phi, \sigma_{h}) = (0.99, 0.10)$ -- the Exp prior again leads to excessive penalization, even when $\sigma_{\alpha} = 0.10$, the highest value considered. Therefore, the Exp prior demonstrates inconsistent performance across scenarios, suggesting its inadequacy for the DynSSV-t model.

\begin{table}[h!]
    \caption{Results of $\hat{\bm{\theta}}$ estimates of PCP model for different values of $\sigma_{\alpha}$ based on $m=300$ replicates of $T=1500$. True value $\mu=0.0$ and $\nu=8.0$.}
    \centering
    \scalebox{1.0}{
    \begin{tabular}{l c rrrrrrr}
    \toprule
    $\sigma_{\alpha}$ &$\bm{\theta}$ & Mean & Inf & Sup & CD & $\text{Bias}_{rel}$ & $\text{RMSE}_{rel}$ & Coverage \\  
    \midrule
    & \multicolumn{8}{c}{$(\phi,\sigma_{h})=(0.95, 0.15)$} \\
    \cmidrule(l){2-9}
    \multirow{5}{*}{0.01}     &$\mu$               &0.007  &-0.188  &0.202  &0.07  &0.007  &0.096  &0.94 \\
                                                     &$\phi$              &0.947  &0.903  &0.984  &0.04  &-0.003 &0.021  &0.97 \\
                                                     &$\sigma_{h}$        &0.140  &0.080  &0.209  &-0.04 &-0.066 &0.221  &0.95 \\
                                                     &$\sigma_{\alpha}$   &0.021  &0.001  &0.050  &0.05  &1.052  &1.502  &0.98 \\
                                                     &$\nu$               &10.400 &5.680  &17.236 &-0.11 &0.300  &0.535  &0.96 \\
    \addlinespace
    \multirow{5}{*}{0.05}     &$\mu$               &0.002  &-0.191  &0.194  &-0.02 &0.002  &0.094  &0.93 \\
                                                     &$\phi$              &0.948  &0.904  &0.983  &-0.05 &-0.003 &0.018  &0.96 \\
                                                     &$\sigma_{h}$        &0.140  &0.081  &0.207  &0.05  &-0.070 &0.200  &0.94 \\
                                                     &$\sigma_{\alpha}$   &0.059  &0.018  &0.110  &0.05  &0.172  &0.563  &0.90 \\
                                                     &$\nu$               &10.065 &5.805  &15.966 &0.03  &0.258  &0.515  &0.94 \\
    \addlinespace
    \multirow{5}{*}{0.10}     &$\mu$               &0.007  &-0.195  &0.223  &-0.05 &0.007  &0.113  &0.95 \\
                                                     &$\phi$              &0.949  &0.906  &0.984  &0.02  &-0.002 &0.020  &0.94 \\
                                                     &$\sigma_{h}$        &0.138  &0.081  &0.203  &0.03  &-0.080 &0.196  &0.96 \\
                                                     &$\sigma_{\alpha}$   &0.121  &0.050  &0.205  &-0.02 &0.205  &0.534  &0.91 \\
                                                     &$\nu$               &9.795  &6.033  &14.780 &0.06  &0.224  &0.426  &0.93 \\

    \cmidrule(l){2-9}
    & \multicolumn{8}{c}{$(\phi,\sigma_{h})=(0.99, 0.10)$} \\
    \cmidrule(l){2-9}
    \multirow{5}{*}{0.01}    &$\mu$              &-0.011 &-0.578 &0.553  &0.01  &-0.011 &0.246  &0.93 \\
                            &$\phi$              &0.985  &0.971  &0.997  &0.06  &-0.005 &0.010  &0.95 \\
                            &$\sigma_{h}$        &0.107  &0.072  &0.147  &-0.00 &0.073  &0.179  &0.99 \\
                            &$\sigma_{\alpha}$   &0.018  &0.001  &0.044  &-0.06 &0.837  &1.335  &0.98 \\
                            &$\nu$               &10.323 &5.828  &16.739 &-0.14 &0.290  &0.501  &0.95 \\
    \addlinespace
    \multirow{5}{*}{0.05}   &$\mu$               &0.037  &-0.495  &0.576  &0.09  &0.037  &0.217  &0.94 \\
                            &$\phi$              &0.984  &0.971  &0.996  &0.00  &-0.006 &0.010  &0.96 \\
                            &$\sigma_{h}$        &0.110  &0.074  &0.149  &0.08  &0.095  &0.199  &0.96 \\
                            &$\sigma_{\alpha}$   &0.056  &0.019  &0.101  &-0.08 &0.117  &0.467  &0.94 \\
                            &$\nu$               &10.429 &6.088  &16.494 &-0.00 &0.304  &0.516  &0.94 \\
    \addlinespace
    \multirow{5}{*}{0.10}   &$\mu$               &0.065  &-0.459  &0.579  &0.04  &0.065  &0.238  &0.92 \\
                            &$\phi$              &0.984  &0.971  &0.996  &-0.00 &-0.006 &0.010  &0.95 \\
                            &$\sigma_{h}$        &0.108  &0.073  &0.146  &0.02  &0.076  &0.176  &0.97 \\
                            &$\sigma_{\alpha}$   &0.117  &0.053  &0.193  &-0.10 &0.174  &0.460  &0.93 \\
                            &$\nu$               &10.077 &6.130  &15.641 &0.02  &0.260  &0.454  &0.94 \\
    \bottomrule
    \end{tabular}
    \label{tab3}
    }
\end{table}

Finally, Table \ref{tab3} presents the results obtained using the PCP prior. Overall, the estimates of the parameter $\sigma_{\alpha}$ exhibit consistently good performance across different settings, characterized by relatively low relative bias and RMSE values, along with high coverage rates, even in more demanding scenarios. In the case of $(\phi, \sigma_{h}) = (0.95, 0.15)$, it is observed that, even for $\sigma_{\alpha}=0.01$, showing minimal bias and a coverage rate near $98\%$, which indicates high reliability in the quantification of uncertainty. As $\sigma_{\alpha}$ increases, the performance remains satisfactory, with reductions in the RMSE values and maintenance of good coverage. Under the most persistent regime, with $(\phi, \sigma_{h}) = (0.99, 0.10)$, the estimates of $\sigma_{\alpha}$ continue to stand out. Even at $\sigma_{\alpha}=0.01$, the error metrics are compatible with scenarios of greater variability, and the coverage remains above $93\%$ in all cases. 
\begin{figure}[h!]
     \centering
     \begin{subfigure}[b]{1.0\textwidth}
         \centering
         \includegraphics[width=1.0\textwidth]{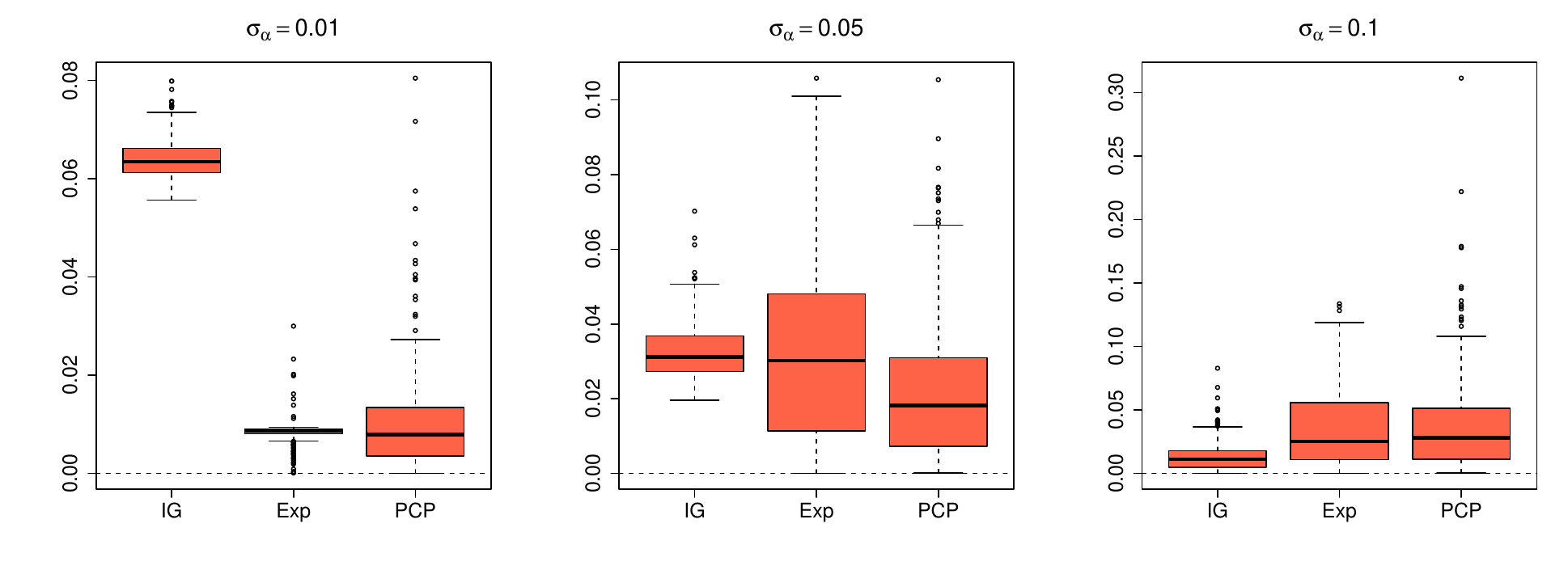}
         \caption{$(\phi, \sigma_{h}) = (0.95, 0.15)$.}
     \end{subfigure}
     \vspace{0.1cm}\\
     \begin{subfigure}[b]{1.0\textwidth}
         \centering
         \includegraphics[width=1.0\textwidth]{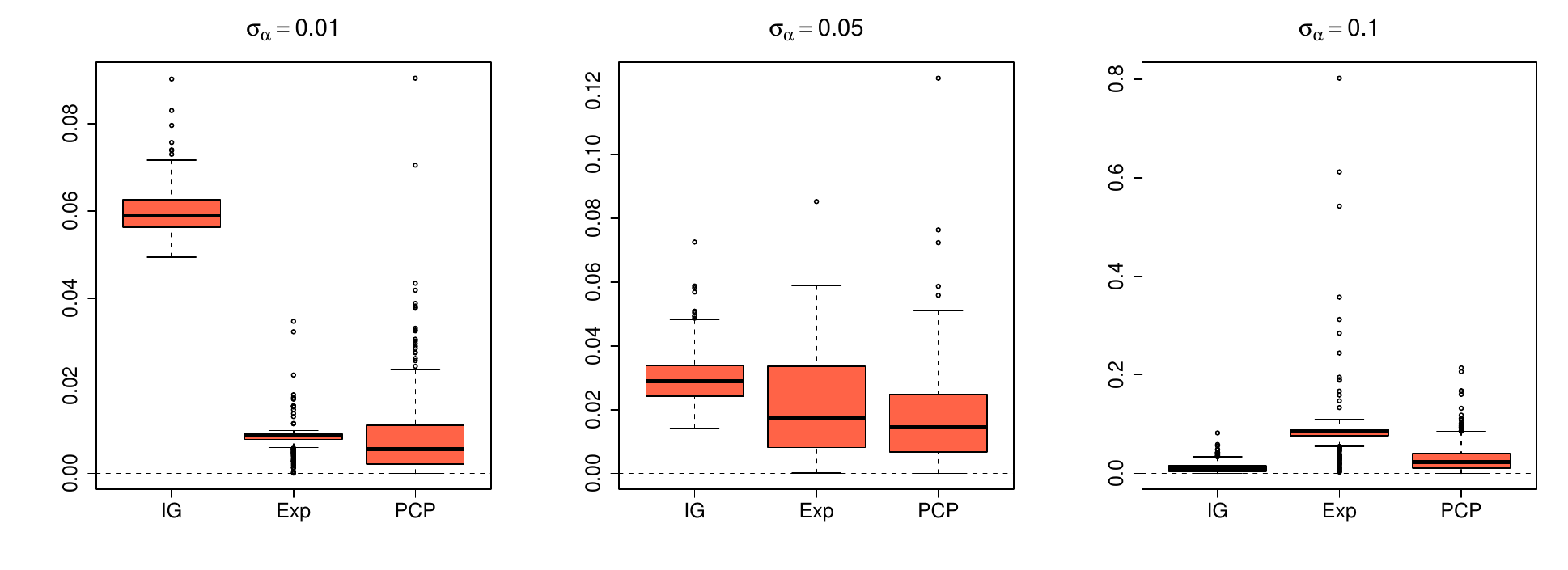}
         \caption{$(\phi, \sigma_{h}) = (0.99, 0.10)$.}
     \end{subfigure}
     \caption{Boxplots of absolute errors $|\hat{\sigma}_{\alpha} - \sigma_{\alpha}|$ of pointwise estimation under different simulation scenarios. The horizontal dashed line indicates the zero-error reference.}
     \label{boxplotssim}
\end{figure}
For $\sigma_{\alpha}=0.10$, the relative bias and RMSE decrease, confirming the robustness of the PCP prior even in environments with high persistence and low variance. These findings indicate that the PCP prior is well calibrated for estimating $\sigma_{\alpha}$ in the DynSSV-t model, providing stable and reliable inference across various persistence and volatility levels.

The strong performance of the PCP prior is again evident in Figure \ref{boxplotssim}, which displays the absolute errors of the point estimates, $|\hat{\sigma}_{\alpha} - \sigma_{\alpha}|$. A systematic pattern is observed for the $\mathcal{IG}$ prior, which tends to introduce excessive complexity in low-variability scenarios ($\sigma_{\alpha} = 0.01, 0.05$), but yields the best performance when $\sigma_{\alpha} \approx 0.10$, which is close to its mode. In contrast, the Exp prior exhibits inconsistent behavior in the scenarios with $\sigma_{\alpha}=0.1$. Therefore, based on these results, the PCP prior was selected for the empirical application.

\section{Real Data Application}
\label{section:Section5}

In this section, the proposed methodology is applied to the daily returns of three major cryptocurrencies, namely Bitcoin, Ethereum, and XRP. Data were obtained from Yahoo Finance (\url{http://finance.yahoo.com}) considering the period from January 7, 2017 to December 6, 2022. The same period for Bitcoin returns was also studied by \cite{atance2024time}. Returns are computed as mean corrected compounded percentage returns, defined as
\begin{equation*}
    y_t = 100 \times \left \{ \log P_t - \log P_{t-1} - \frac{1}{T} \sum_{j=1}^{T} (\log P_j - \log P_{j-1}) \right \},
\end{equation*}
where $P_t$ denotes the closing price on day $t$, for $t=1, ..., T$.

\begin{table}[h!]
    \caption{Summary statistics of the return indexes.}
    \centering
    \begin{tabular}{l c c c c c c c}
    \toprule
    Return & T    & Mean & SD & Min. & Max. & Skewness & Kurtosis  \\
    \midrule
    Bitcoin  & 1982     & 0.00  & 4.15  & -46.64    & 22.34     & -0.69     & 13.54 \\
    Ethereum & 1676     &0.00   & 5.17  &-55.16     & 23.38     & -0.96     & 13.09 \\
    XRP      & 1676     &0.00   & 6.57  &-55.08     & 60.66     & 0.83      & 18.94 \\
    \bottomrule
    \label{statistics}
    \end{tabular}
\end{table}

Table \ref{statistics} reports descriptive statistics for the mean-corrected compounded returns. For each return series, we compute the mean, standard deviation, skewness, and kurtosis. All series exhibit considerable variability, with standard deviations ranging from 4.15 to 6.57. Notably, the XRP index displays the widest range, with a minimum value of -55.08 and a maximum close to 60. Additionally, the kurtosis values are substantially greater than 3, and the presence of marked skewness suggests that daily returns deviate from the assumption of symmetric normality. To account for these features, we analyze the data using the DynSSV-SMSN class models introduced in Section \ref{section:Section2}, which offers a more flexible and robust framework. For comparison purposes, we also estimate the static versions of these models (StatSSV-SMSN) to better illustrate the role of dynamic asymmetry.

For all models, we conducted the MCMC simulation for 52000 iterations, discarding the first 50000 as warm-up. The results found are based on the last 2000 iterations and are organized in Tables \ref{bitcoin}, \ref{ethereum} and \ref{xrp}. According to the CD values, the null hypothesis that the sequence of 2000 draws is stationary is accepted at the $5\%$ level for all parameters and in all the models considered, since the DC statistics are contained in the range $(-1.96, 1.96)$. We also note that all inefficiency factors (IF) are close to 1, indicating good performance of the samples.

The posterior mean estimates for $\mu$, which represents the level of log-volatility $h_{t}$, were found around 2.3 and 3.5, and are statistically significant in all models, indicating a considerable level of volatilities. The estimated values of $(\phi, \sigma_{h})$ are consistent with previous findings in the literature (see \cite{abanto2010robust}), except for the SN models. This discrepancy can be attributed to the structural limitation of SN models in capturing volatility dynamics under the presence of outliers, as the model is unable to distinguish between extreme observations and genuine changes in latent volatility. As a result, the model tends to overestimate $\sigma_{h}$, leading to a downward bias in the estimate of $\phi$, relative to tail-heavy models. 

\begin{table}[h!]
    \caption{Estimation results for the Bitcoin returns. First row: Posterior mean. Second row: HDP $95\%$ credible interval. Third row: CD statistics. Fourth row: Inefficient factor.}
    \centering
    \scalebox{0.9}{
    \begin{tabular}{l c c c c c c }
    \toprule
    Parameter                       & StatSSV-N             & DynSSV-N                & StatSSV-t         & DynSSV-t          & StatSSV-S       & DynSSV-S  \\
    \midrule
    \multirow{4}{*}{$\mu$}          & 2.30                 & 2.29                    &  2.56            & 2.61             & 2.86            & 2.83 \\
                                    & (2.11, 2.52)         & (2.06, 2.52)            & (2.10, 2.97)     & (2.14, 3.05)     & (2.30, 3.48)    & (2.29, 3.41) \\
                                    & 1.07                 & -0.74                    & 0.40             & -1.21            & -0.65           & -0.16 \\
                                    & 1.00                 & 1.00                     & 1.00             & 1.00             & 1.00            & 0.97 \\
    \addlinespace
    \multirow{4}{*}{$\phi$}         & 0.88                 & 0.88                     &  0.98            & 0.98             & 0.98            & 0.98 \\
                                    & (0.83, 0.93)         & (0.83, 0.93)            & (0.96, 0.99)     & (0.96, 0.99)     & (0.96, 0.99)    & (0.96, 0.99) \\
                                    & 0.29                 & -1.44                    & 0.34             & -0.40            & 0.80            & 1.20 \\
                                    & 1.00                 & 1.08                     & 1.03             & 1.10             & 1.00            & 1.00 \\
    \addlinespace
    \multirow{4}{*}{$\sigma_{h}$}   & 0.52                 & 0.51                     &  0.20            & 0.19             & 0.20            & 0.20 \\
                                    & (0.41, 0.63)         & (0.39, 0.63)            & (0.15, 0.25)     & (0.14, 0.25)     & (0.15, 0.26)    & (0.16, 0.25) \\
                                    & -0.93                & 1.12                     & -1.76            & 0.39             & 0.13            & -1.18 \\
                                    & 1.00                 & 1.00                     & 1.00             & 1.00             & 1.00            & 1.00 \\
    \addlinespace
    \multirow{4}{*}{$\alpha_{1}$}   & 0.02                 & 0.13                     &  0.02            & 0.18             & 0.01            & 0.21 \\
                                    & (-0.06, 0.09)        & (-0.12, 0.48)           & (-0.06, 0.09)    & (-0.13, 0.57)    & (-0.06, 0.09)   & (-0.14, 0.59) \\
                                    & -0.94                & 0.55                     & 1.20             & -0.41            & 1.40            & 1.23 \\
                                    & 1.00                 & 1.01                     & 1.00             & 1.00             & 1.00            & 0.92 \\
    \addlinespace
    \multirow{4}{*}{$\sigma_{\alpha}$}   & -               & 0.02                     &  -               & 0.02             & -               & 0.02 \\
                                         & -               & (0.00, 0.04)             & -               & (0.00, 0.04)     & -               & (0.00, 0.04) \\
                                         & -               & -1.01                     & -               & 0.39             & -               & -0.06 \\
                                         & -               & 1.00                      & -               & 1.00             & -               & 1.00 \\
    \addlinespace
    \multirow{4}{*}{$\nu$}          & -                  & -                        &  4.12            & 3.79             & 1.33            & 1.34 \\
                                    & -                  & -                        & (3.25, 5.00)     & (3.02, 4.59)     & (1.13, 1.54)    & (1.15, 1.55) \\
                                    & -                  & -                        & 1.87             & 0.62             & 1.04            & 0.60 \\
                                    & -                  & -                        & 1.07             & 1.00             & 0.96            & 0.94 \\
    \bottomrule
    \end{tabular}
    }
    \label{bitcoin}
\end{table}

\begin{table}[h!]
    \caption{Estimation results for the Ethereum returns. First row: Posterior mean. Second row: HDP $95\%$ credible interval. Third row: CD statistics. Fourth row: Inefficient factor.}
    \centering
    \scalebox{0.9}{
    \begin{tabular}{l c c c c c c }
    \toprule
    Parameter                       & StatSSV-N             & DynSSV-N              & StatSSV-t           & DynSSV-t           & StatSSV-S         & DynSSV-S  \\
    \midrule
    \multirow{4}{*}{$\mu$}          & 2.82                 & 2.83                  & 3.03                & 3.06                & 3.19              & 3.16 \\
                                    & (2.66, 3.02)         & (2.67, 3.00)          & (2.73, 3.33)        & (2.76, 3.32)        & (2.84, 3.55)      & (2.81, 3.54) \\
                                    & 0.10                 & 0.58                  & -0.04               & -0.41               & 0.61              & 1.16 \\
                                    & 1.00                 & 1.08                  & 1.00                & 1.00                & 1.00              & 0.86 \\
    \addlinespace
    \multirow{4}{*}{$\phi$}         & 0.85                 & 0.84                  & 0.96                & 0.96                & 0.96              & 0.96 \\
                                    & (0.78, 0.91)         & (0.76, 0.91)          & (0.92, 0.98)        & (0.93, 0.98)        & (0.93, 0.98)      & (0.93, 0.98) \\
                                    & 1.51                 & 0.43                  & 1.66                & 1.59                & 0.86              & -0.38 \\
                                    & 1.00                 & 1.08                  & 1.00                & 1.00                & 1.00              & 1.00 \\
    \addlinespace
    \multirow{4}{*}{$\sigma_{h}$}   & 0.49                 & 0.52                  & 0.21                & 0.20                & 0.20              & 0.21 \\
                                    & (0.37, 0.61)         & (0.39, 0.63)          & (0.12, 0.29)        & (0.13, 0.29)        & (0.13, 0.29)      & (0.13, 0.29) \\
                                    & -1.64                & -0.97                 & -1.17               & -1.10               & -1.24             & -0.42 \\
                                    & 1.15                 & 1.08                  & 1.09                & 1.00                & 1.00              & 1.00 \\
    \addlinespace
    \multirow{4}{*}{$\alpha_{1}$}   & 0.02                 & 0.02                  & 0.02                & 0.06                & 0.02              & 0.06 \\
                                    & (-0.06, 0.09)        & (-0.04, 0.09)         & (-0.05, 0.09)       & (-0.18, 0.35)       & (-0.05, 0.09)     & (-0.16, 0.37) \\
                                    & -0.33                & -1.19                 & 0.05                & -0.28               & 0.40              & -0.60 \\
                                    & 1.00                 & 1.00                  & 1.00                & 1.07                & 1.20              & 1.00 \\
    \addlinespace
    \multirow{4}{*}{$\sigma_{\alpha}$}   & -                 & 0.01                  & -                   & 0.01                & -                 & 0.01 \\
                                         & -                 & (0.00, 0.03)          & -                   & (0.00, 0.03)        & -                 & (0.00, 0.03) \\
                                         & -                 & -1.69                 & -                   & -0.07               & -                 & 0.44 \\
                                         & -                 & 1.07                  & -                   & 1.00                & -                 & 1.00 \\
    \addlinespace
    \multirow{4}{*}{$\nu$}          & -                     & -                     & 5.65                & 5.10                & 1.57              & 1.60 \\
                                    & -                     & -                     & (4.03, 7.33)        & (3.60, 7.52)        & (1.26, 1.92)      & (1.28, 1.92) \\
                                    & -                     & -                     & -2.23               & -0.96               & -0.71             & 0.01 \\
                                    & -                     & -                     & 1.00                & 1.00                & 1.00              & 1.00 \\
    \bottomrule
    \end{tabular}
    }
    \label{ethereum}
\end{table}

\begin{table}[h!]
    \centering
    \caption{Estimation results for the XRP returns. First row: Posterior mean. Second row: HDP $95\%$ credible interval. Third row: CD statistics. Fourth row: Inefficient factor.}
    \scalebox{0.9}{
        \begin{tabular}{l c c c c c c}
        \toprule
        Parameter               & StatSSV-N         & DynSSV-N          & StatSSV-t         & DynSSV-t          & StatSSV-S         & DynSSV-S \\
        \midrule
        \multirow{4}{*}{$\mu$}  & 2.73             & 2.73              & 3.07              & 3.19              & 3.53              & 3.50 \\
                                & (2.50, 3.00)     & (2.48, 2.96)      & (2.64, 3.50)      & (2.70, 3.66)      & (2.80, 4.39)      & (2.85, 4.23) \\
                                & -1.27            & 0.87              & -1.49             & -0.18             & -1.26             & 1.49 \\
                                & 1.00             & 1.00              & 1.00              & 1.24              & 1.00              & 1.00 \\
        \addlinespace
        \multirow{4}{*}{$\phi$} & 0.84             & 0.85              & 0.95              & 0.96              & 0.96              & 0.96 \\
                                & (0.79, 0.89)     & (0.80, 0.90)      & (0.92, 0.98)      & (0.94, 0.99)      & (0.93, 0.98)      & (0.93, 0.98) \\
                                & 1.79             & -0.14             & -1.43             & 0.62              & 0.59              & -0.22 \\
                                & 1.00             & 1.00              & 1.00              & 1.21              & 1.00              & 1.00 \\
        \addlinespace
        \multirow{4}{*}{$\sigma_h$} & 0.74          & 0.73              & 0.34              & 0.29              & 0.32              & 0.32 \\
                                    & (0.63, 0.86)  & (0.62, 0.87)      & (0.21, 0.48)      & (0.19, 0.40)      & (0.20, 0.45)      & (0.21, 0.43) \\
                                    & -1.75         & -1.14             & 1.92              & 0.08              & -0.79             & -0.25 \\
                                    & 1.00          & 1.00              & 1.23              & 0.96              & 1.00              & 1.00 \\
        \addlinespace
        \multirow{4}{*}{$\alpha_1$} & -0.04         & -0.02             & -0.04             & -0.02             & -0.04             & -0.02 \\
                                    & (-0.12, 0.04) & (-0.23, 0.13)     & (-0.12, 0.03)     & (-0.21, 0.16)     & (-0.12, 0.03)     & (-0.19, 0.17) \\
                                    & -1.10         & -0.10             & 0.00              & -0.02             & 0.48              & 0.97 \\
                                    & 1.00          & 0.92              & 0.88              & 1.00              & 1.00              & 1.00 \\
        \addlinespace
        \multirow{4}{*}{$\sigma_{\alpha}$}  & -         & 0.01              & -                 & 0.01              & -                 & 0.01 \\
                                            & -         & (0.00, 0.02)     & -                 & (0.00, 0.02)     & -                 & (0.00, 0.02) \\
                                            & -         & 0.14              & -                 & -1.45             & -                 & 1.40 \\
                                            & -         & 1.00              & -                 & 1.08              & -                 & 1.30 \\
        \addlinespace
        \multirow{4}{*}{$\nu$}  & -             & -                 & 4.40              & 3.59              & 1.29              & 1.29 \\
                                & -             & -                 & (2.66, 6.44)      & (2.88, 4.53)      & (1.03, 1.56)      & (1.07, 1.53) \\
                                & -             & -                 & 1.25              & 0.26              & 0.77              & -0.95 \\
                                & -             & -                 & 1.11              & 1.09              & 1.00              & 1.00 \\
        \bottomrule
    \end{tabular}
    }
    \label{xrp}
\end{table}

The posterior mean of $\alpha_1$ varies substantially across different model specifications, likely due to the increased complexity of the dynamic models. This complexity is reflected in the wider HPD intervals observed in the models with time-varying skewness (DynSSV), compared to the narrower intervals in static specifications. In the static models, the estimates of $\alpha_1$ tend to cluster around zero, suggesting a negligible effect of skewness when it is assumed to be constant over time. In contrast, dynamic models yield higher posterior means for $\alpha_1$, which may indicate an improved ability to capture conditional asymmetry in returns. 

Regarding the $\sigma_{\alpha}$ parameter, the posterior means were approximately 0.02 for Bitcoin returns, 0.01 for Ethereum, and 0.006 for XRP, suggesting a relatively modest degree of variability. Moreover, the lower bounds of the HPD intervals include zero for both Ethereum and XRP, but not for Bitcoin. This provides stronger evidence of time-varying asymmetry in the case of Bitcoin. Moreover, these results reinforce the fact that the PCP is capable of accommodating different levels of asymmetry variability, as also observed in simulation studies.

The parameter $\nu$ governs the heaviness of the distribution’s tails. For both Bitcoin and XRP returns, the posterior means of $\nu$ were approximately 4 in the Student's t models and around 1.3 in the Slash models. For Ethereum, the corresponding means were slightly higher, around 5 for the t model and 1.5 for the Slash model — all values close to the lower bounds of their respective domains ($\nu > 2$ for Student-t and $\nu > 1$ for Slash). These results indicate the presence of extremely heavy tails in cryptocurrency returns and further highlight the inadequacy of models with normal errors for capturing volatility dynamics in such markets, as previously discussed.

For model selection, we employ the Deviance Information Criterion (DIC), the Watanabe-Akaike Information Criterion (WAIC), and Leave-One-Out Cross-Validation (LOO-CV). In all cases, lower values indicate better model fit. According to Table \ref{criterios}, the DynSSV-t model exhibited the best performance across all three criteria. It is also worth noting that models assuming normality yielded the highest values for the information criteria, further highlighting their poor fit under this assumption. Furthermore, we note that dynamic models present better fits than static ones.

\begin{table}[h!]
    \caption{Calculated information criteria for Bitcoin, Ethereum and XRP cryptocurrencies.}
    \centering
    \begin{tabular}{l l c c c }
    \toprule
    Aplication & Model           & DIC         & WAIC        & LOO-CV       \\
    \midrule
    Bitcoin &StatSSV-N       & 10394.19    & 10502.00    & 10572.71     \\
            &DynSSV-N        & 10382.68    & 10492.69    & 10573.49      \\
            &StatSSV-t       & 10225.24    & 10296.94    & 10403.48     \\
            &DynSSV-t        & \textbf{10190.38}    & \textbf{10266.02}    & \textbf{10382.07}     \\
            &StatSSV-S       & 10311.65    & 10317.65    & 10422.64     \\
            &DynSSV-S        & 10291.68    & 10302.26    & 10415.08     \\
    \midrule
    Ethereum &StatSSV-N       & 9695.55          & 9761.94          &9811.88     \\
             &DynSSV-N        & 9681.36          & 9750.51          &9805.45      \\
             &StatSSV-t       & 9626.04          & 9675.77          &9748.14     \\
             &DynSSV-t        & \textbf{9606.44} & \textbf{9655.14} & \textbf{9732.04}     \\
             &StatSSV-S       & 9686.37          & 9689.83          &9761.75     \\
             &DynSSV-S        & 9674.98          & 9681.90          &9756.17     \\
    \midrule
    XRP     &StatSSV-N       & 9559.32          & 9662.92            & 9749.04     \\
            &DynSSV-N        & 9564.69          & 9670.29            & 9761.00      \\
            &StatSSV-t       & 9538.49          & 9623.40            & 9724.33     \\
            &DynSSV-t        & \textbf{9527.52} & \textbf{9609.10}   & \textbf{9717.28}     \\
            &StatSSV-S       & 9597.79          & 9632.78            & 9740.77     \\
            &DynSSV-S        & 9596.43          & 9632.28            & 9740.75     \\
    \bottomrule
    \end{tabular}
    \label{criterios}
\end{table}

\begin{figure}[h!]
    \centering
    \includegraphics[width=0.95\textwidth]{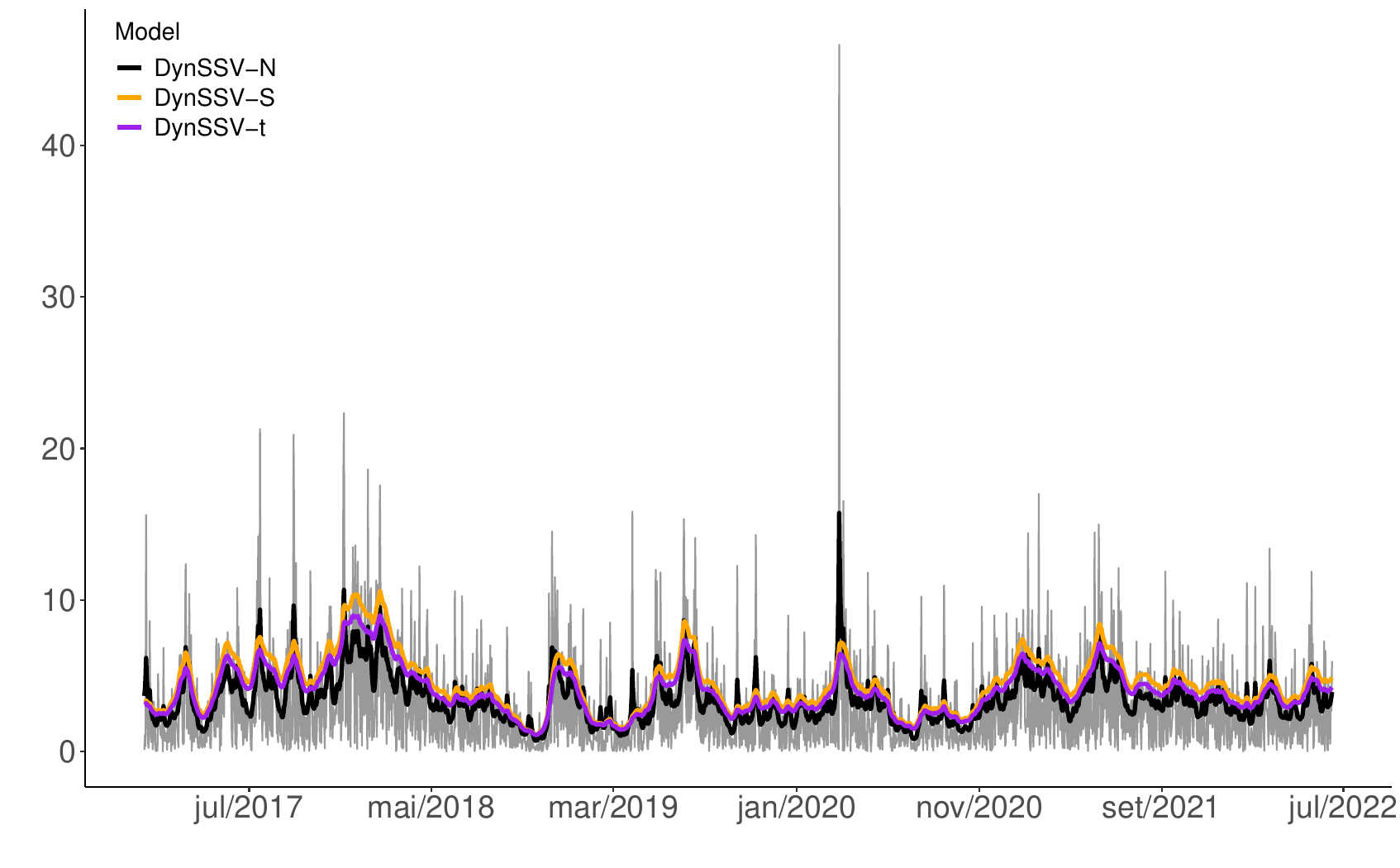}
    \caption{Absolute returns (gray line) of the Bitcoin indices and the corresponding $exp\{h_t/2\}$ estimates derived from the DynSSV-N (black lines), DynSSV-t (purple line), and DynSSV-S (orange line) models.}
    \label{vol_bitcoin}
\end{figure}

\begin{figure}[h!]
    \centering
    \includegraphics[width=0.95\textwidth]{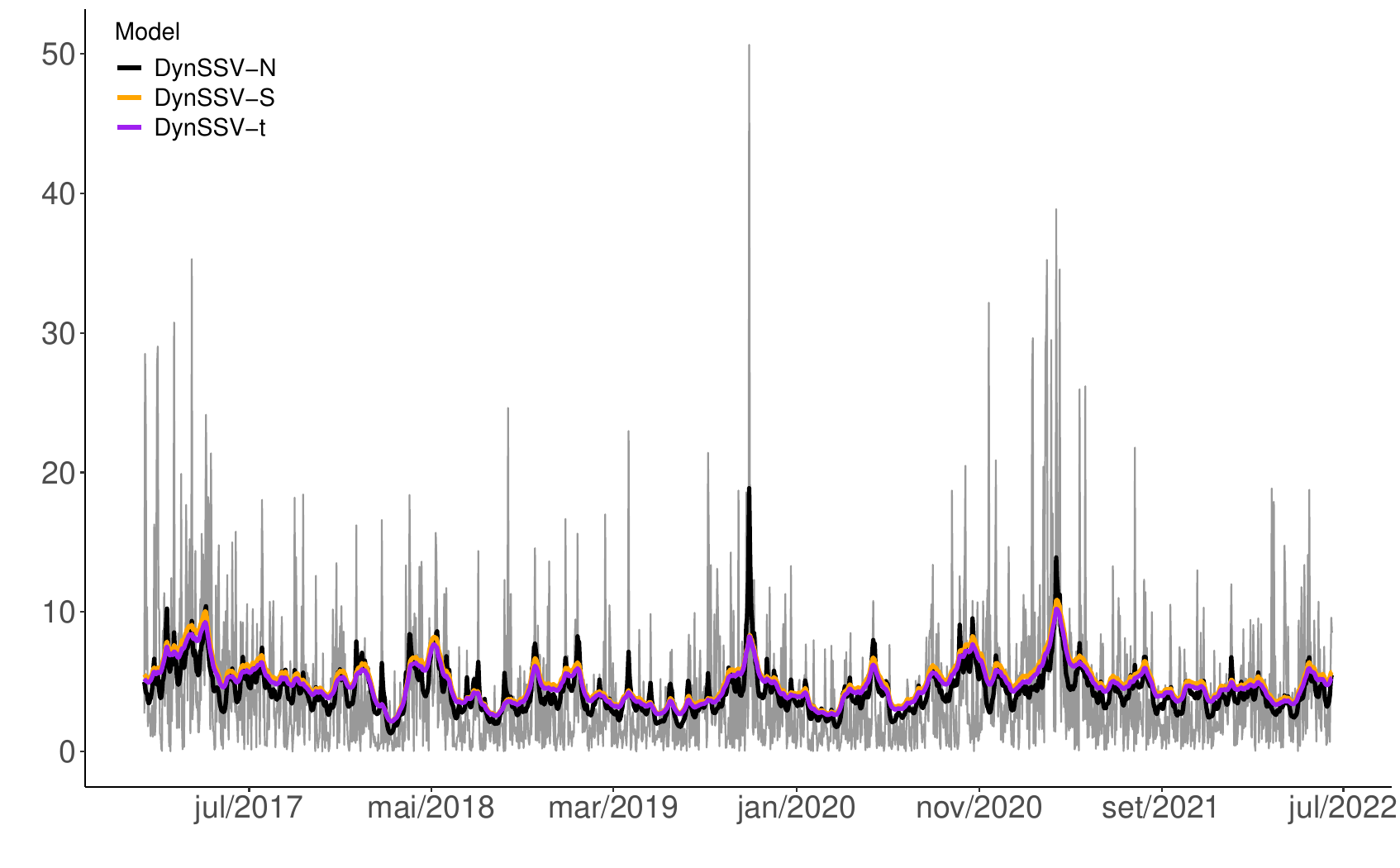}
    \caption{Absolute returns (gray line) of the Ethereum indices and the corresponding $exp\{h_t/2\}$ estimates derived from the DynSSV-N (black lines), DynSSV-t (purple line), and DynSSV-S (orange line) models.}
    \label{vol_ethereum}
\end{figure}

\begin{figure}[h!]
    \centering
    \includegraphics[width=0.95\textwidth]{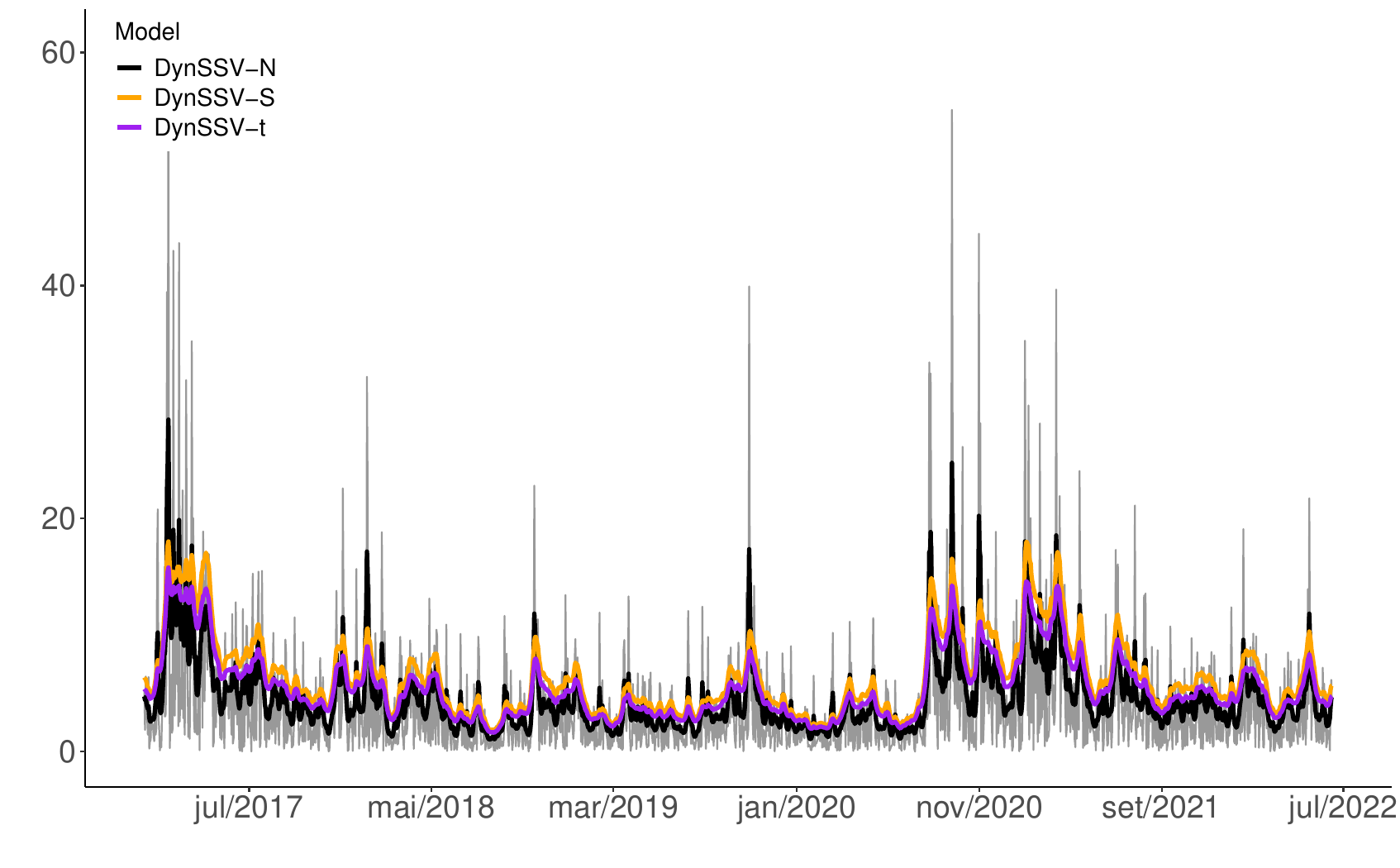}
    \caption{Absolute returns (gray line) of the XRP indices and the corresponding $exp\{h_t/2\}$ estimates derived from the DynSSV-N (black lines), DynSSV-t (purple line), and DynSSV-S (orange line) models.}
    \label{vol_xrp}
\end{figure}

Figures \ref{vol_bitcoin}, \ref{vol_ethereum} and \ref{vol_xrp} displays the estimated volatilities, $\exp\{\hat{h}_t/2\}$, in contrast to the absolute returns, where $\hat{h}_t$ denotes the posterior mean of the latent log-volatility. Overall, the estimated volatilities closely follow the periods of highest variability in the return series. Notably, the DynSSV-t and DynSSV-S models exhibit smoother dynamics compared to the DynSSV-N model. Furthermore, the extreme observation associated with the onset of the COVID-19 crisis, around May 2020, has a pronounced effect on the DynSSV-N volatility estimates. The smoother volatility patterns produced by the DynSSV-t and DynSSV-S models suggest that accounting for heavy tails leads to a more stable and realistic characterization of market uncertainty. In particular, during episodes of financial stress, such as the COVID-19 shock, models that fail to accommodate non-Gaussian features tend to overreact, as seen in the spiked estimates of the DynSSV-N model. This behavior can distort risk assessments and misguide policy or investment decisions. This further underscores the robustness of the DynSSV-SMSN class model structures in handling outliers.

\begin{figure}[h!]
    \centering

    \begin{subfigure}[t]{\textwidth}
        \centering
        \begin{subfigure}[t]{0.3\textwidth}
            \centering
            \includegraphics[width=\textwidth]{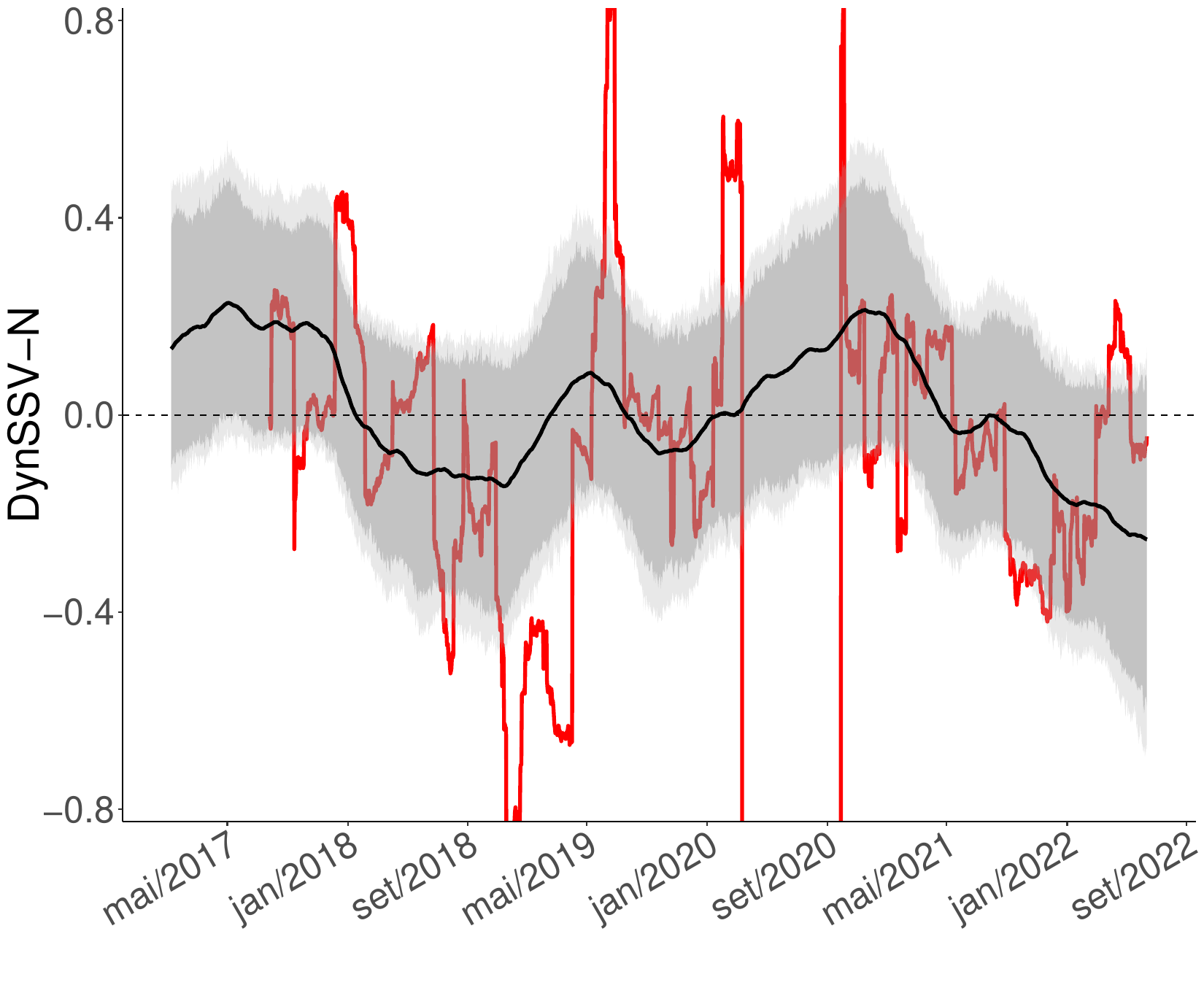}
        \end{subfigure}
        \hfill
        \begin{subfigure}[t]{0.3\textwidth}
            \centering
            \includegraphics[width=\textwidth]{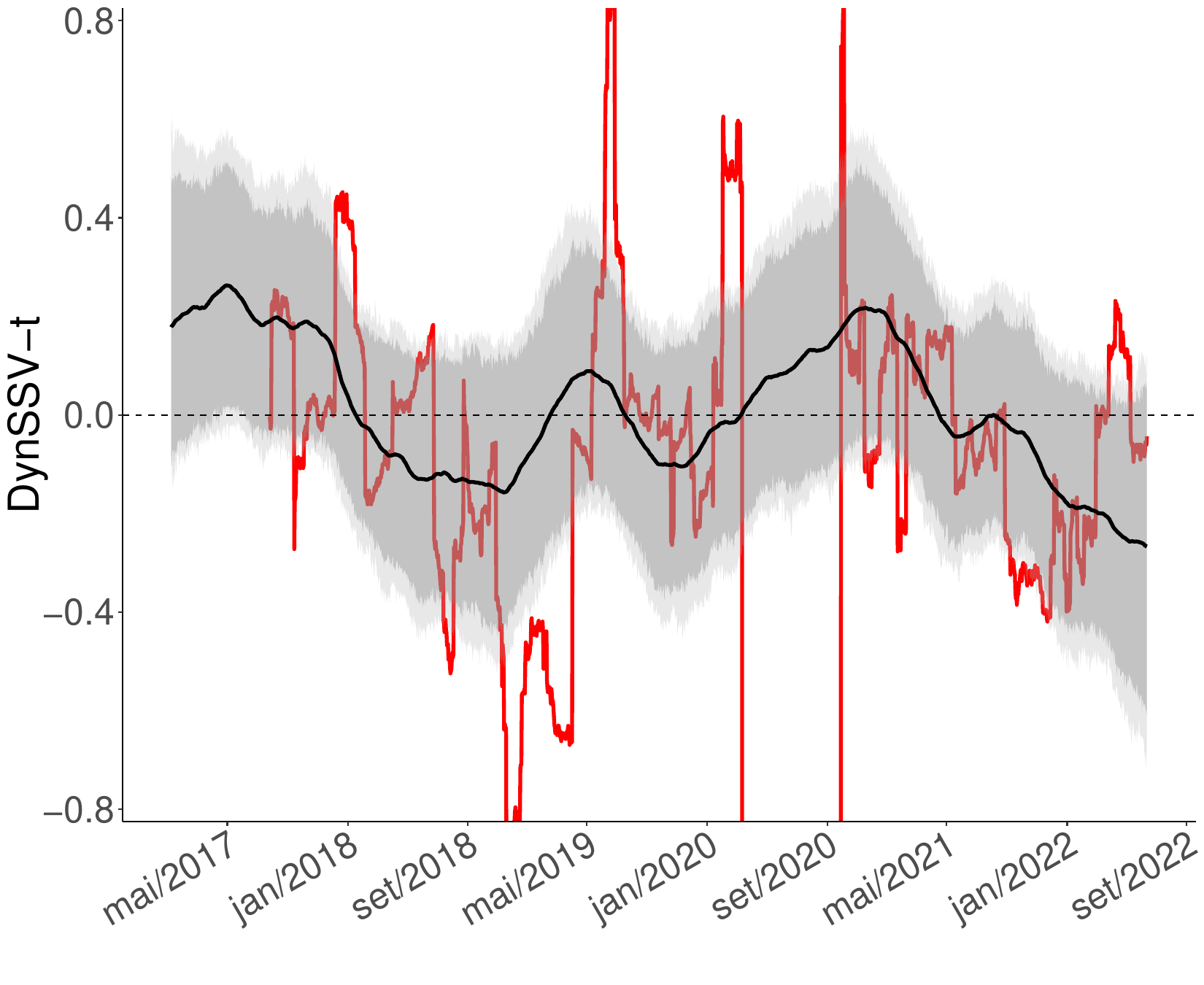}
        \end{subfigure}
        \hfill
        \begin{subfigure}[t]{0.3\textwidth}
            \centering
            \includegraphics[width=\textwidth]{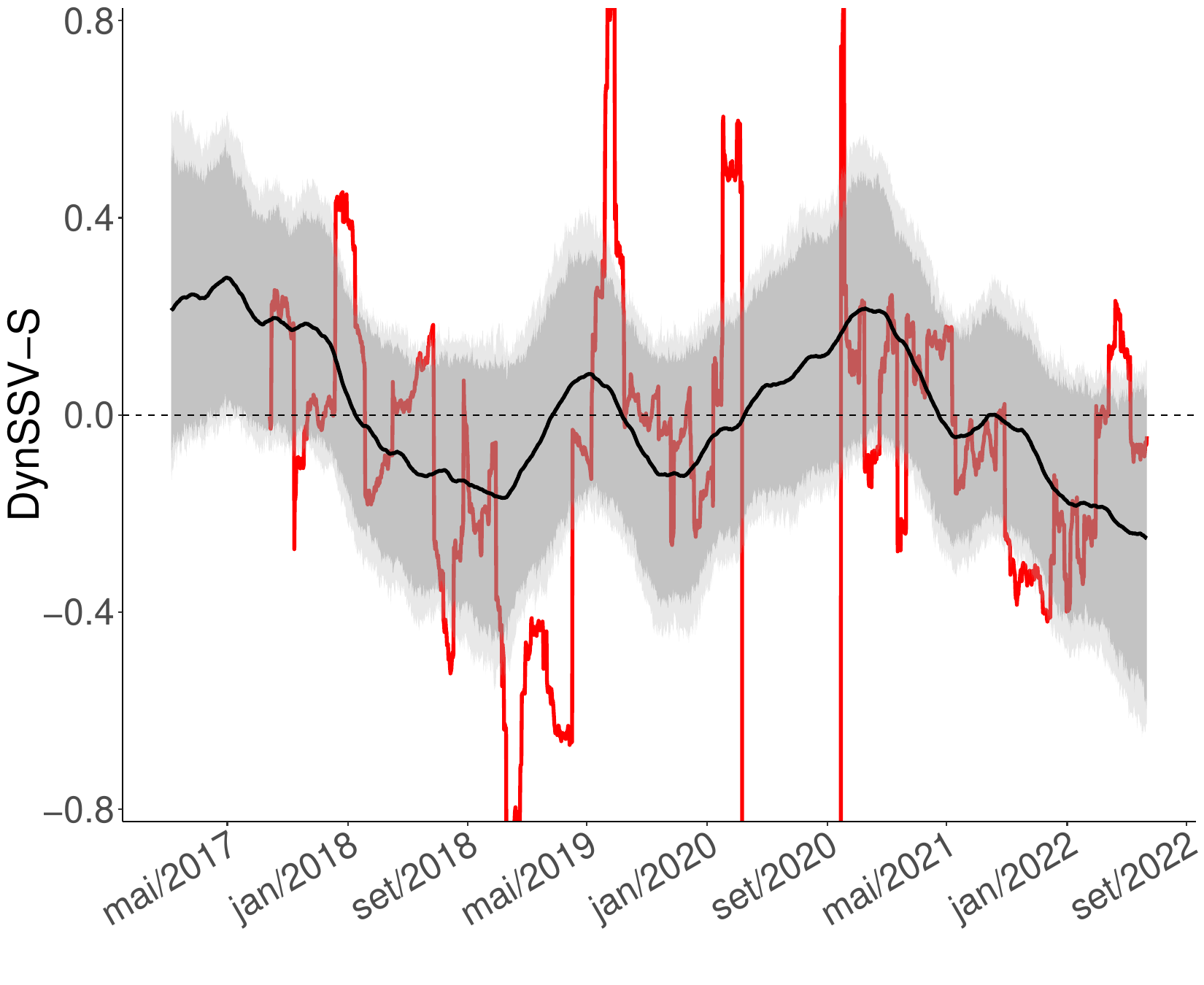}
        \end{subfigure}
        \caption*{\textbf{(a)} Bitcoin returns}
    \end{subfigure}

    \vspace{0.4cm}

    \begin{subfigure}[t]{\textwidth}
        \centering
        \begin{subfigure}[t]{0.3\textwidth}
            \centering
            \includegraphics[width=\textwidth]{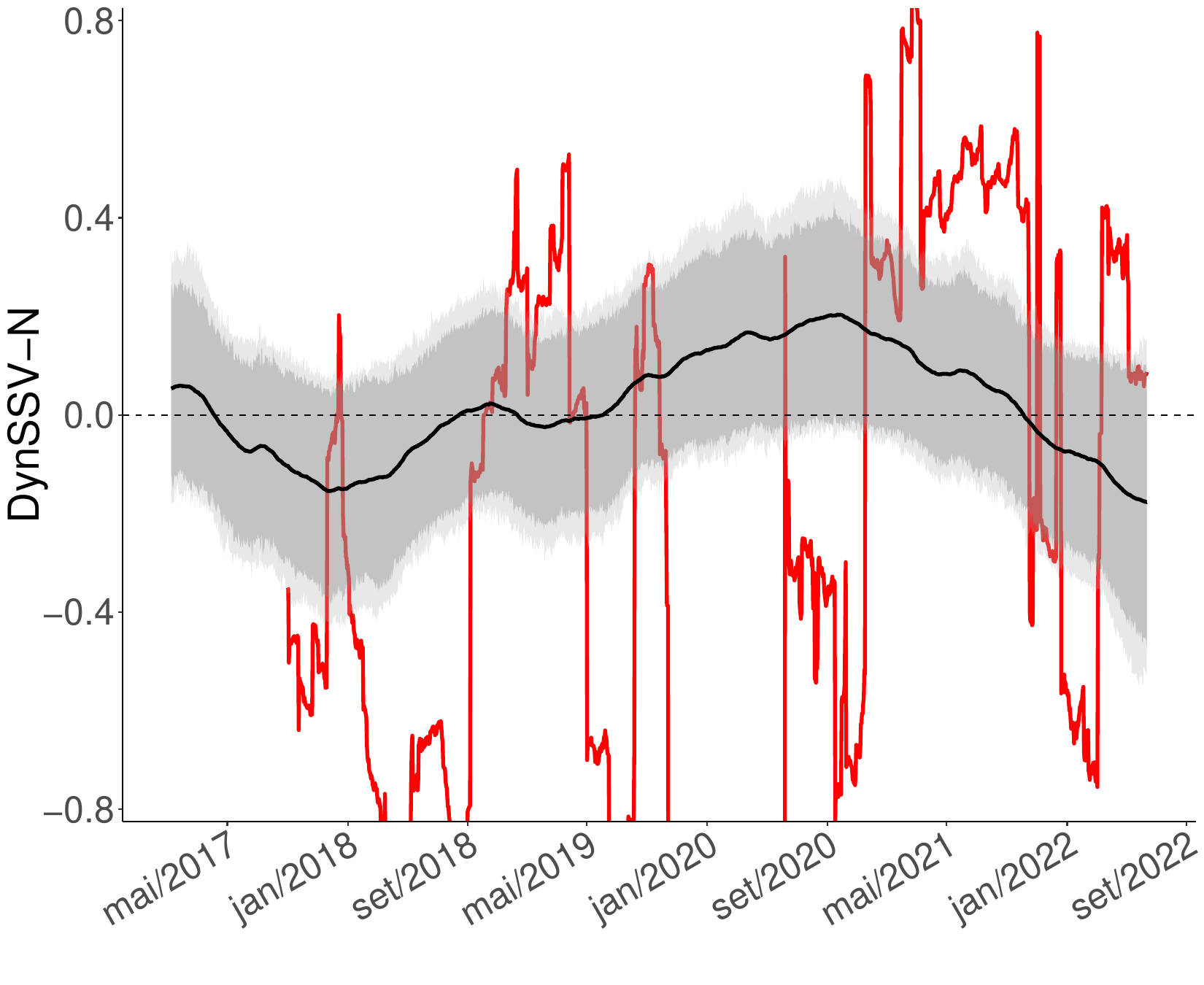}
        \end{subfigure}
        \hfill
        \begin{subfigure}[t]{0.3\textwidth}
            \centering
            \includegraphics[width=\textwidth]{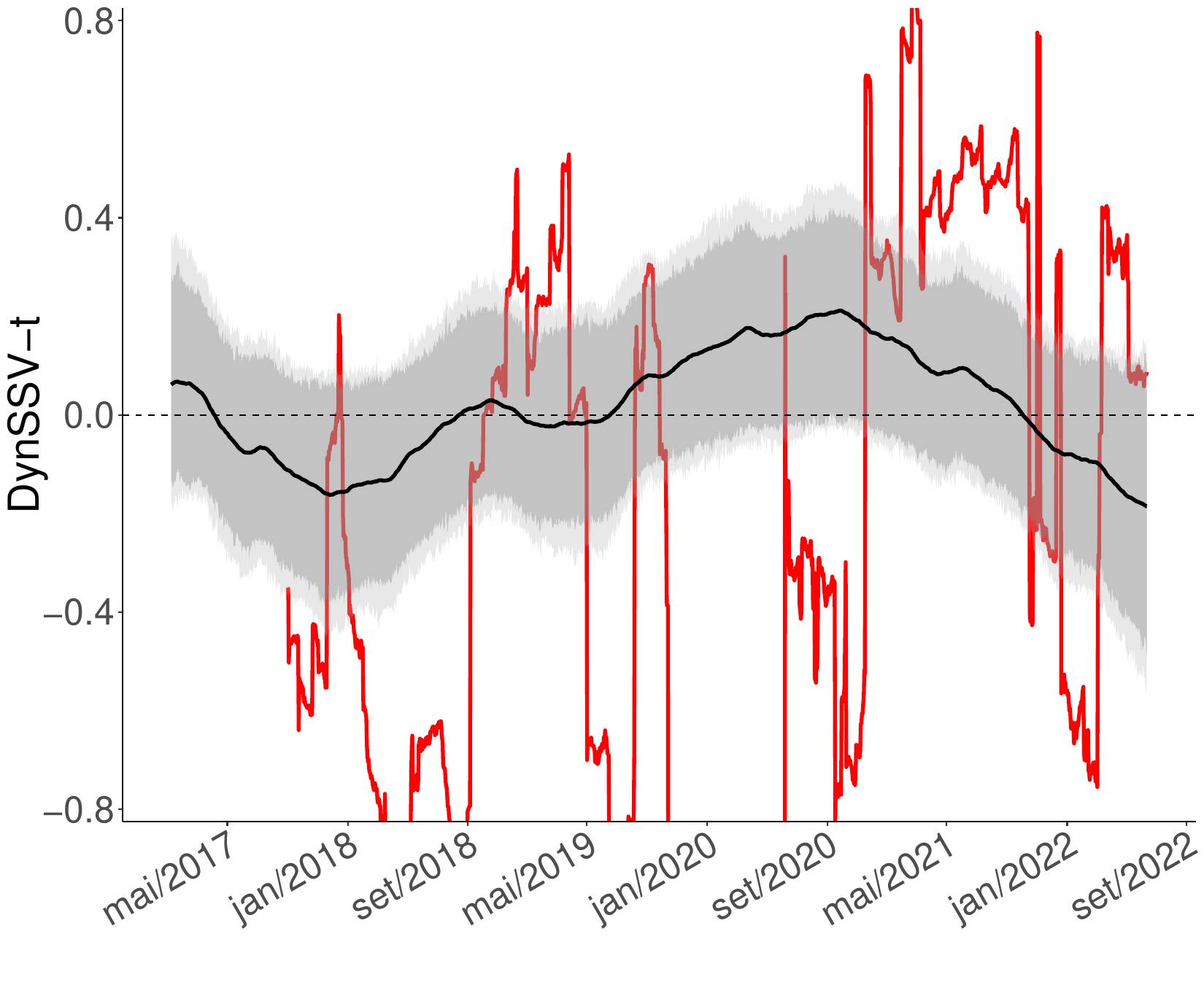}
        \end{subfigure}
        \hfill
        \begin{subfigure}[t]{0.3\textwidth}
            \centering
            \includegraphics[width=\textwidth]{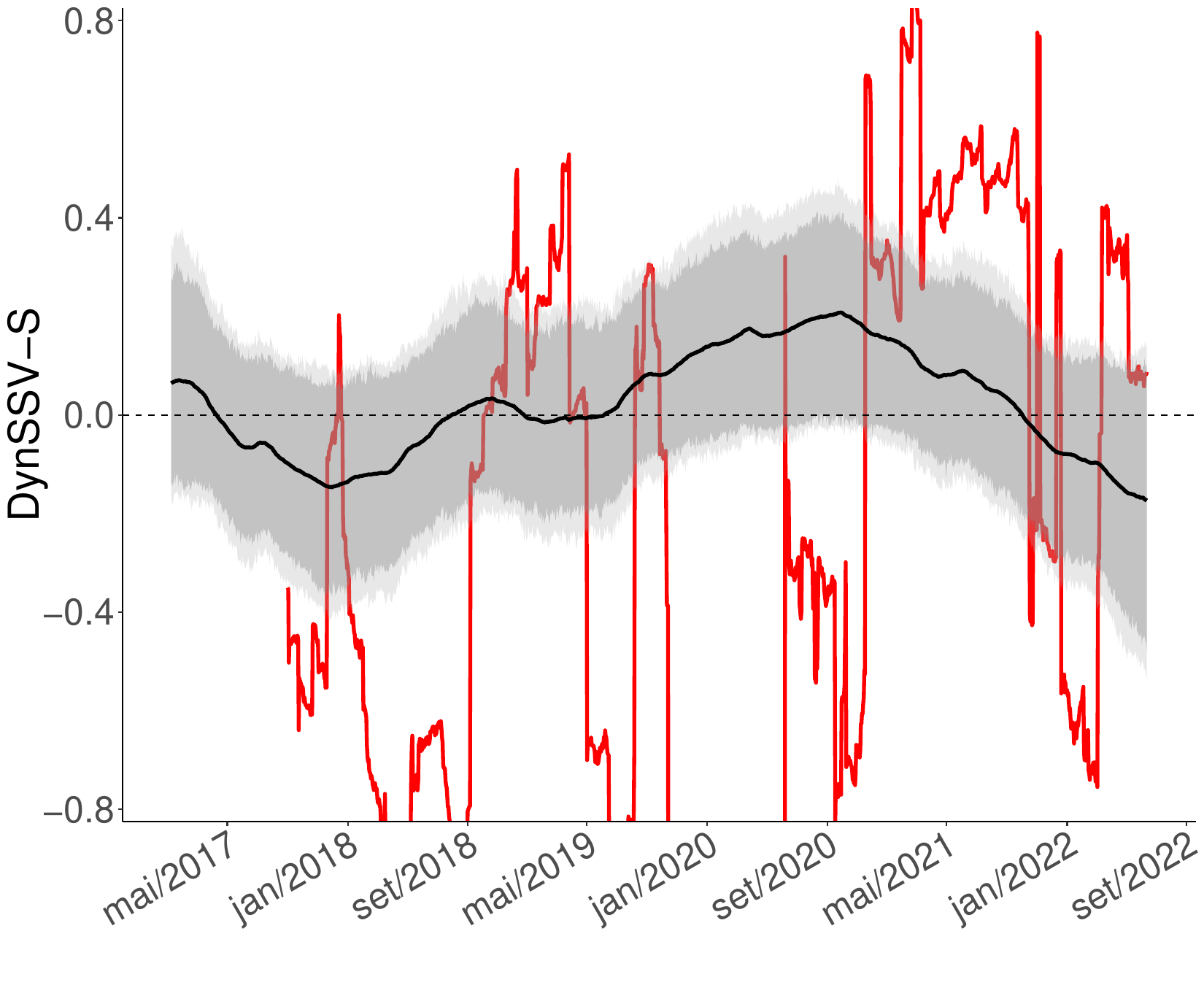}
        \end{subfigure}
        \caption*{\textbf{(b)} Ethereum returns}
    \end{subfigure}

    \vspace{0.4cm}

    \begin{subfigure}[t]{\textwidth}
        \centering
        \begin{subfigure}[t]{0.3\textwidth}
            \centering
            \includegraphics[width=\textwidth]{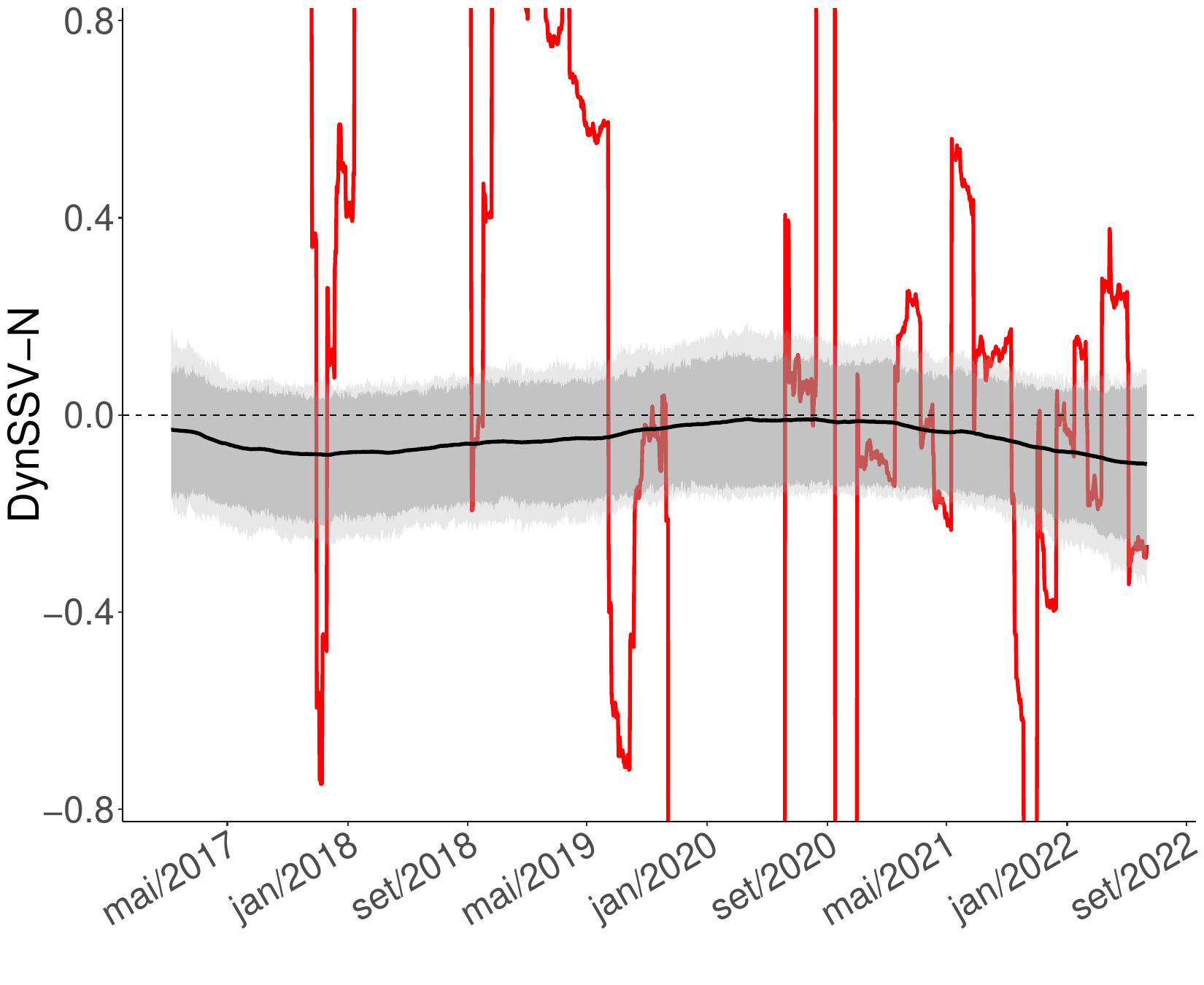}
        \end{subfigure}
        \hfill
        \begin{subfigure}[t]{0.3\textwidth}
            \centering
            \includegraphics[width=\textwidth]{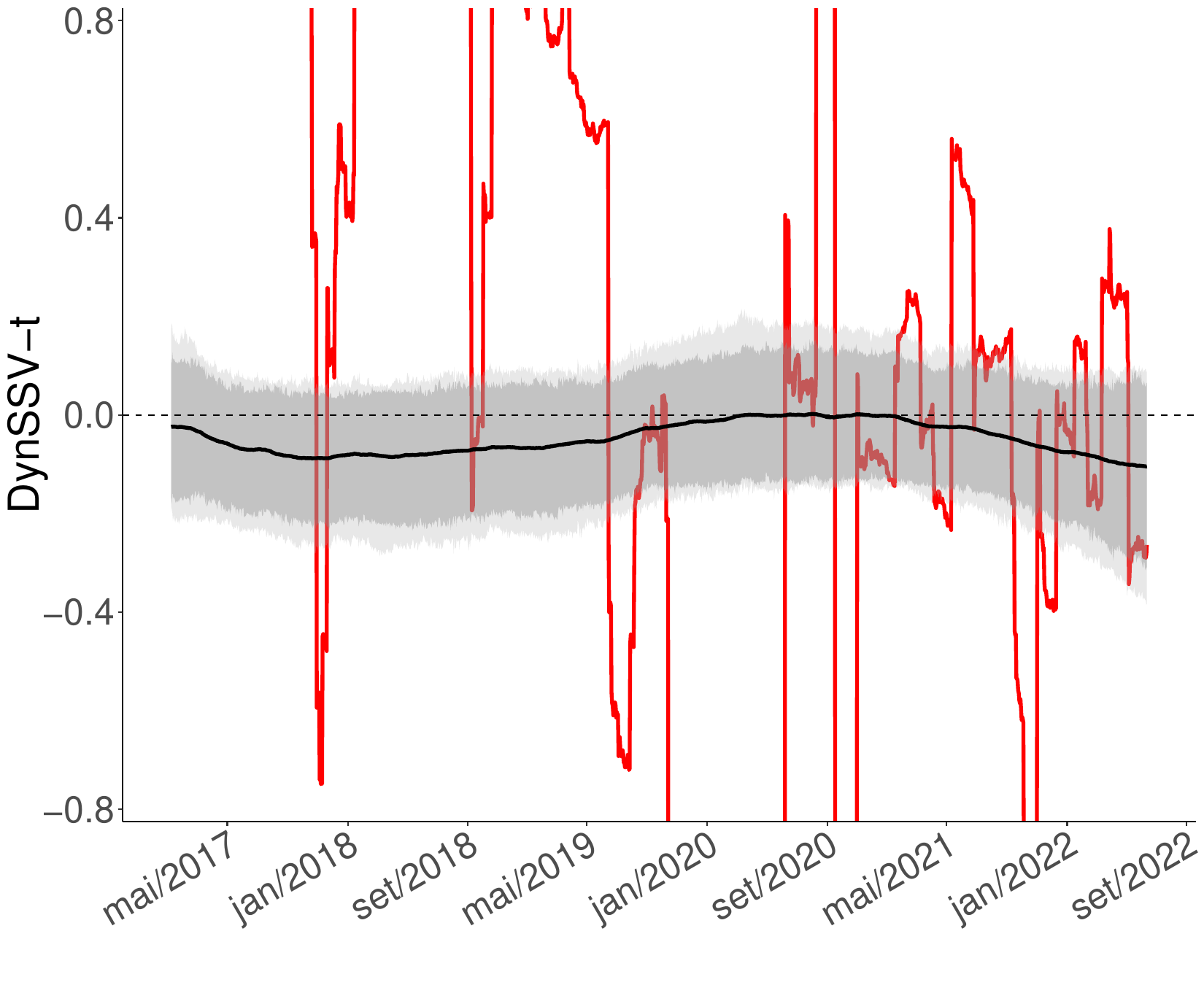}
        \end{subfigure}
        \hfill
        \begin{subfigure}[t]{0.3\textwidth}
            \centering
            \includegraphics[width=\textwidth]{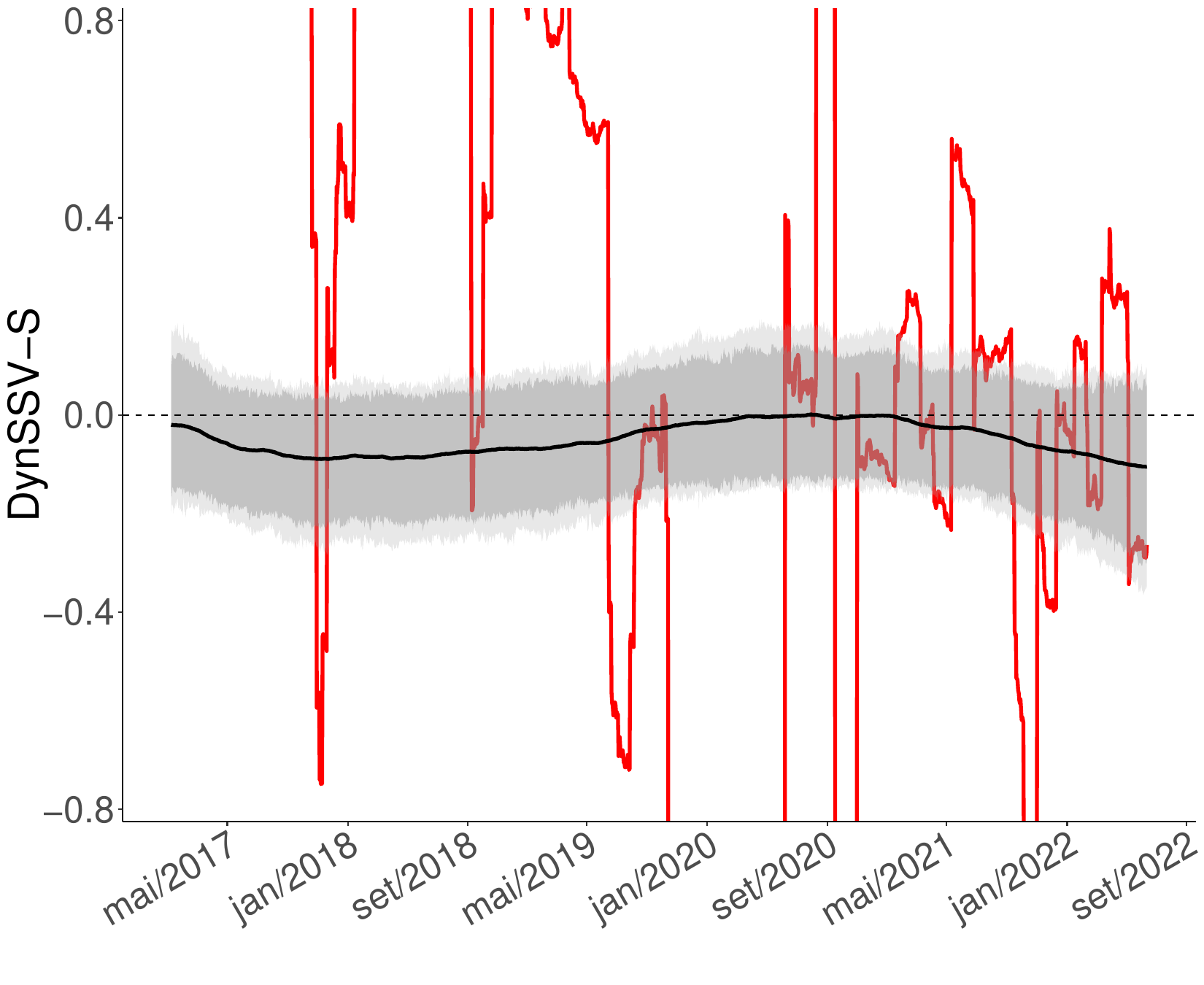}
        \end{subfigure}
        \caption*{\textbf{(c)} XRP returns}
    \end{subfigure}

    \caption{Estimated conditional skewness \textit{versus} empirical skewness. The black line represents the posterior mean of $\hat{\alpha}_t$, while the dark (light) shaded area denotes the HDP $90\%$ ($95\%$) credible interval. The red line corresponds to the empirical skewness.}
    \label{skew}
\end{figure}

To evaluate the ability of the proposed models to capture time-varying asymmetry in returns, we compare the estimated conditional skewness with the empirical unconditional skewness computed on a rolling window fashion of 200 observations, following the approach of \cite{atance2024time}. Figures \ref{skew} display the posterior mean of the time-varying skewness parameter (black line), along with its credible HDP $90\%$ ($95\%$) intervals (shaded areas) and the corresponding empirical skewness (red line) for each model. 

For Bitcoin returns, the estimated skewness tends to follow major changes in empirical skewness for all three models. In particular, empirical skewness is affected by a sharp decline around May 2020, coinciding with the onset of the COVID-19 crisis, which led to an intensification of the volatility process (Figure \ref{vol_bitcoin}). However, the estimated conditional skewness presents greater stability and smoother dynamics in the skewness parameter over time, which is in line with its robustness observed in previous analyses. These results reinforce the relevance of incorporating heavy tails and skewness in modeling financial time series with sudden changes in distributional form. Furthermore, it is worth noting that although the HPD 95\% intervals include zero for all models throughout the analyzed period, this is not the case for the HPD 90\% interval of the DynSSV-t and DynSSV-S models. As shown in Figure \ref{skew}, skewness is positive and statistically significant during a period around May 2017. These findings highlight the effectiveness of models that allow for dynamic skewness in capturing the evolving distributional features of Bitcoin returns. 

In contrast, for Ethereum and XRP returns, the estimated skewness exhibits smoother behavior than their respective empirical skewness measures, indicating more stable distributional dynamics. Additionally, for both assets, the 90\% and 95\% HPD intervals consistently include zero throughout the entire period, suggesting that there is no strong statistical evidence of persistent or time-varying asymmetry. This result illustrates the inherent challenge of estimating skewness accurately \cite{iseringhausen2020time} in financial time series, particularly when the data provide limited information due to the low signal-to-noise ratio typical of such settings \cite{bitto2019achieving}.

\section{Conclusion}
\label{section:Section6}

In this paper, we develop a class of stochastic volatility models with a time-varying skewness parameter, using the Skewed Scale Mixtures of Normal (SMSN) family of distributions. This framework allows for the joint modeling of skewness and heavy tails in return distributions. The proposed class includes specifications with Student-\textit{t}, Slash, and Normal innovations.

To control model complexity, we incorporated priors that penalize deviations from the simpler case with static skewness. A classical prior frequently found in the literature was also considered for comparison. Through simulation studies, we showed that the Penalized Complexity Prior (PCP) consistently outperforms the classical alternative by reducing the risk of overfitting, especially in scenarios with weak evidence of skewness dynamics.

Based on these findings, we adopt the above PCP in empirical application to the returns of major cryptocurrencies. The results indicate that models accommodating both dynamic skewness and heavy tails offer a superior fit to the data, as measured by in-sample information criteria. In particular, we observed substantial variation in asymmetry over time, with positively skewed returns at the beginning of the sample period under the Student-\textit{t} and Slash specifications for Bitcoin returns.

Future directions for this work include evaluating the predictive performance of DynSSV-SMSN models in terms of risk measures such as Value at Risk (VaR). To mitigate the computational burden associated with VaR estimation, recent advances in MCMC techniques, such as the approach proposed by \cite{yang2017value}, can be explored. In addition, the DynSSV-SMSN framework can be extended to incorporate exogenous covariates into the dynamic equation for the skewness parameter, as in \cite{renzetti2023modelling}, potentially improving its explanatory power.

All computational routines developed in this study (\texttt{.R} and \texttt{.stan} files) are available in the GitHub repository at the following link: 
\url{https://github.com/holtz27/dynssv-smsn}.

\section*{Acknowledgments}
 The authors would like to thank the EULER cluster team of the CeMEAI project (FAPESP grant 2013/07375-0) for their support and availability, which enabled the computational simulations and analyses conducted in this study. The first author also thanks Professor Josemar Rodrigues for the valuable discussions that contributed to the development of this work.

\section*{Disclosure statement}
The authors report there are nocompeting interests to declare.

\section*{Data Availability Statement}
The data used in this work were obtained from Yahoo Finance: Bitcoin see \url{https://finance.yahoo.com/quote/BTC-USD/}, Ethereum see \url{https://finance.yahoo.com/quote/ETC-USD/}, and XRP access \url{https://finance.yahoo.com/quote/XRP-USD/}.

\bibliographystyle{unsrtnat}
\bibliography{references}  





\section*{Appendix}

\appendix
\section{Moments of SMSN}
\label{apendiceA}

Let $Z \sim SMSN(\gamma, \omega^{2}, \alpha, \nu)$. Note that $Z | U = u \sim SN(\gamma, u^{-1} \omega^{2}, \alpha)$ and, therefore,
\begin{align*}
    E[Z] &= E[ E[Z|U] ] \\
         &= E\left[ \gamma + U^{-1/2} \sqrt{ \frac{2}{\pi} } \omega \delta \right] \\
         &= \gamma + \sqrt{ \frac{2}{\pi} } \omega \delta E[U^{-1/2}] \\
         &= \gamma + \sqrt{ \frac{2}{\pi} } \omega \delta k_{1}
\end{align*}
and
\begin{align*}
    Var[Z] &= Var[ E[Z|U] ] + E[ Var[Z|U] ] \\
           &= Var\left[ \gamma + U^{-1/2} \sqrt{ \frac{2}{\pi} } \omega \delta \right] + E\left[ U^{-1} \omega^{2}\left(1 - \frac{2}{\pi} \right)\delta^{2} \right] \\
           &= \frac{2}{\pi} \omega^{2} \delta^{2}Var[U^{-1/2}] + \omega^{2}\left(1 - \frac{2}{\pi} \delta^{2} \right) E[U^{-1}] \\
           &= \frac{2}{\pi} \omega^{2} \delta^{2} (k_{2} - k_{1}^{2}) + \omega^{2} \left(1 - \frac{2}{\pi} \delta^{2} \right)k_{2} \\
           &= \omega^{2} \left( k_{2} - \frac{2}{\pi}\delta^{2}k_{1}^{2} \right).
\end{align*}
Particularly, if $Z \sim SN(\gamma, \omega^{2}, \alpha)$, we have $E[Z] = \gamma + \sqrt{ \frac{2}{\pi} } \omega \delta$ and $Var[Z] = \omega^{2} \left(1 - \frac{2}{\pi}\delta^{2} \right)$. That is, in this case $k_{1}=k_{2}=1$.

\section{Hyperparameter Analysis}
\label{apendiceB}

First, note that,
\begin{equation}
    \left( k_{2} - \frac{2}{\pi}\delta_{t}^{2}k_{1}^{2} \right) := \zeta_{t} > 0  \Leftrightarrow  \delta^{2}_{t} < \frac{\pi}{2} \frac{k_{2}}{k_{1}^{2}} := g,
    \label{condition}
\end{equation}
where $g = g(\nu) \in \mathbb{R}$ and $\delta_{t} = \alpha_{t} / \sqrt{1 + \alpha_{t}^{2}} \in (-1, 1), \forall t=1, \ldots, T$. Therefore, ensuring that $g > 1$, we have $\zeta_{t} > 0, \forall \nu \in \mathbb{R}_{+}$. In the Skew normal case, it is easy to see that condition (\ref{condition}) is guaranteed since $\delta^{2}_{t} < \frac{\pi}{2}, \forall \delta_{t} \in (-1,1)$. 

The remainder of this section is intended to show that condition (\ref{condition}) is satisfied for the rest of the models considered in this paper.

\begin{itemize}
    \item \textbf{t-Student}: $U_{t} \sim \mathcal{G}(\nu/2, \nu/2)$.
    \begin{equation*}
        E\left[U^{-m/2}_{t} \right] = \left( \frac{\nu}{2} \right)^{m/2} \frac{\Gamma(\frac{\nu - m}{2})}{\Gamma(\frac{\nu}{2})}
    \end{equation*}
    \begin{equation*}
        k_{1} = \sqrt{ \frac{\nu}{2} } \frac{ \Gamma(\frac{\nu-1}{2}) }{\Gamma(\frac{\nu}{2})} \quad \text{e} \quad k_{2} = \frac{\nu}{\nu - 2}, \quad \nu > 2.
    \end{equation*}
    Then, 
    \begin{equation*}
        g(\nu) = \frac{\pi}{\nu - 2} \frac{\Gamma(\frac{\nu}{2})^{2}}{\Gamma(\frac{\nu-1}{2})^{2}} \overset{\nu \rightarrow \infty}{\longrightarrow} \frac{\pi}{2} > 1.
    \end{equation*}

    \item \textbf{Slash}: $U_{t} \sim \mathcal{B}e(\nu, 1).$
    \begin{equation*}
        E\left[U^{-m/2}_{t}\right] = \frac{2\nu}{2\nu - m}.
    \end{equation*}
    \begin{equation*}
        k_{1} = \frac{2\nu}{2\nu -1} \quad \text{e} \quad k_{2} = \frac{\nu}{\nu -1}, \quad \nu > 1.
    \end{equation*}
    Then, 
    \begin{equation*}
        g(\nu) = \frac{\pi}{8\nu} \frac{(2\nu - 1)^2}{(\nu-1)} \overset{\nu \rightarrow \infty}{\longrightarrow} \frac{\pi}{2} > 1.
    \end{equation*}
\end{itemize}

\end{document}